\documentclass[a4paper]{PoS}
\usepackage[english]{babel}
\usepackage[T1]{fontenc}
\usepackage{graphicx,amsmath,amssymb,bm,bbm}
\usepackage{multirow,booktabs,dcolumn}
\usepackage{isotope,physics}
\usepackage[numbers,sort&compress]{natbib}
\usepackage[capitalize]{cleveref}

\title{Local chiral EFT potentials in nuclei and neutron matter: results and issues}

\ShortTitle{Local chiral EFT potentials}

\author{{\speaker{Diego Lonardoni}$^{\;a,\,b}$ and \speaker{Ingo Tews}$^{\;b}$}\\
     \llap{$^a$}Facility for Rare Isotope Beams, Michigan State University, East Lansing, MI 48824, USA\\
     \llap{$^b$}Theoretical Division, Los Alamos National Laboratory, Los Alamos, NM 87545,\\
     E-mail: \email{lonardoni@nscl.msu.edu}, \email{itews@lanl.gov}
     }

\abstract{In recent years, the combination of advanced quantum Monte Carlo (QMC) methods and local interactions derived from chiral effective field theory (EFT) has been shown to provide a versatile and systematic approach to nuclear systems. Calculations at next-to-next-to-leading order in chiral EFT have lead to fascinating results for nuclei and nucleonic matter. On the one hand, ground-state properties of nuclei are well reproduced up to $A\leq16$, even though these potentials have been fit to nucleon-nucleon scattering and few-body observables only. On the other hand, a reasonable description of neutron-matter properties emerges. While regulator functions applied to two- and three-nucleon forces are a necessary ingredient in these many-body calculations, the use of local regulators leads to a substantial residual regulator and cutoff dependence that increases current theoretical uncertainties. In this contribution, we review local chiral interactions, their applications, and QMC results for nuclei and neutron matter. In addition, we address regulator issues for such potentials and present a possible path forward.}

\FullConference{The 9th International Workshop on Chiral Dynamics\\
		September 17-21, 2018\\
		Durham, NC, USA}

\begin{document}

\makeatletter
\setbox\@firstaubox\hbox{\small D. Lonardoni and I. Tews}
\makeatother

\section{Introduction}

Predicting the emergence of nuclear properties and structure from first principles is a formidable task. A fundamental question is whether it is possible to describe nuclei and their global properties, e.g., binding energies, radii, transitions, and reactions, from microscopic nuclear Hamiltonians fit only to nucleon-nucleon (NN) scattering data and few-body observables, while simultaneously predicting properties of matter, including the equation of state (EOS) and the properties of neutron stars (NS). Despite advanced efforts, definitive answers are not yet available~\cite{Barrett:2013,Hagen:2013,Binder:2013,Lahde:2013,Carlson:2015,Hebeler:2015,Ekstrom:2015,Hergert:2015,Simonis:2017}.

\begin{figure}[b]
\centering
\includegraphics[width=0.31\textwidth]{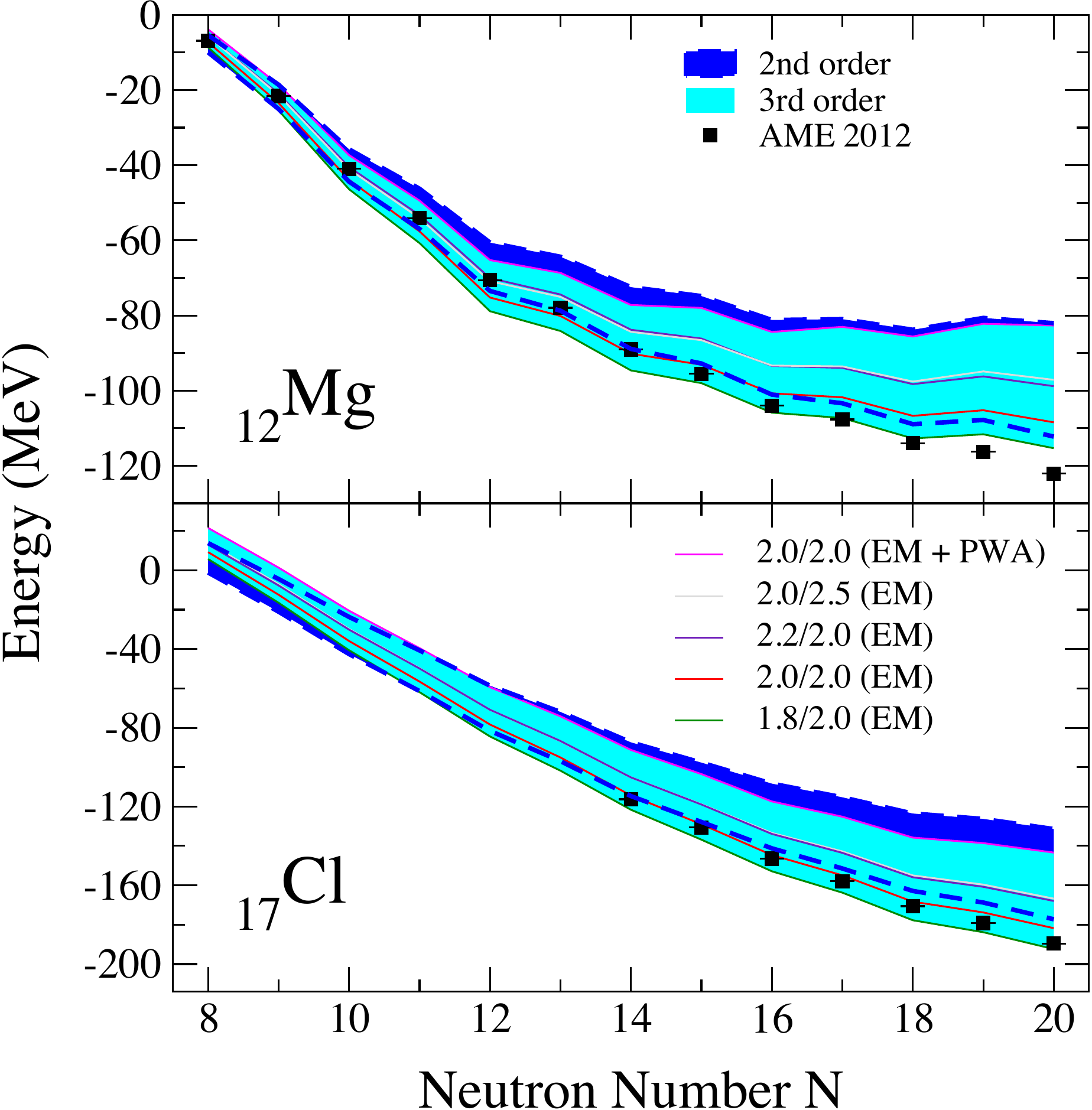}\hfill
\includegraphics[width=.03\textwidth, trim= 0cm 0 18.4cm 17cm, clip=]{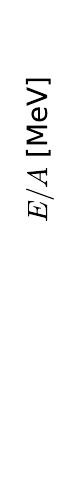}\includegraphics[width=0.30\textwidth, trim= 6.8cm 0.1cm 6.6cm 16cm, clip=]{EOS_panel_ex6.pdf}\hfill
\includegraphics[width=0.31\textwidth]{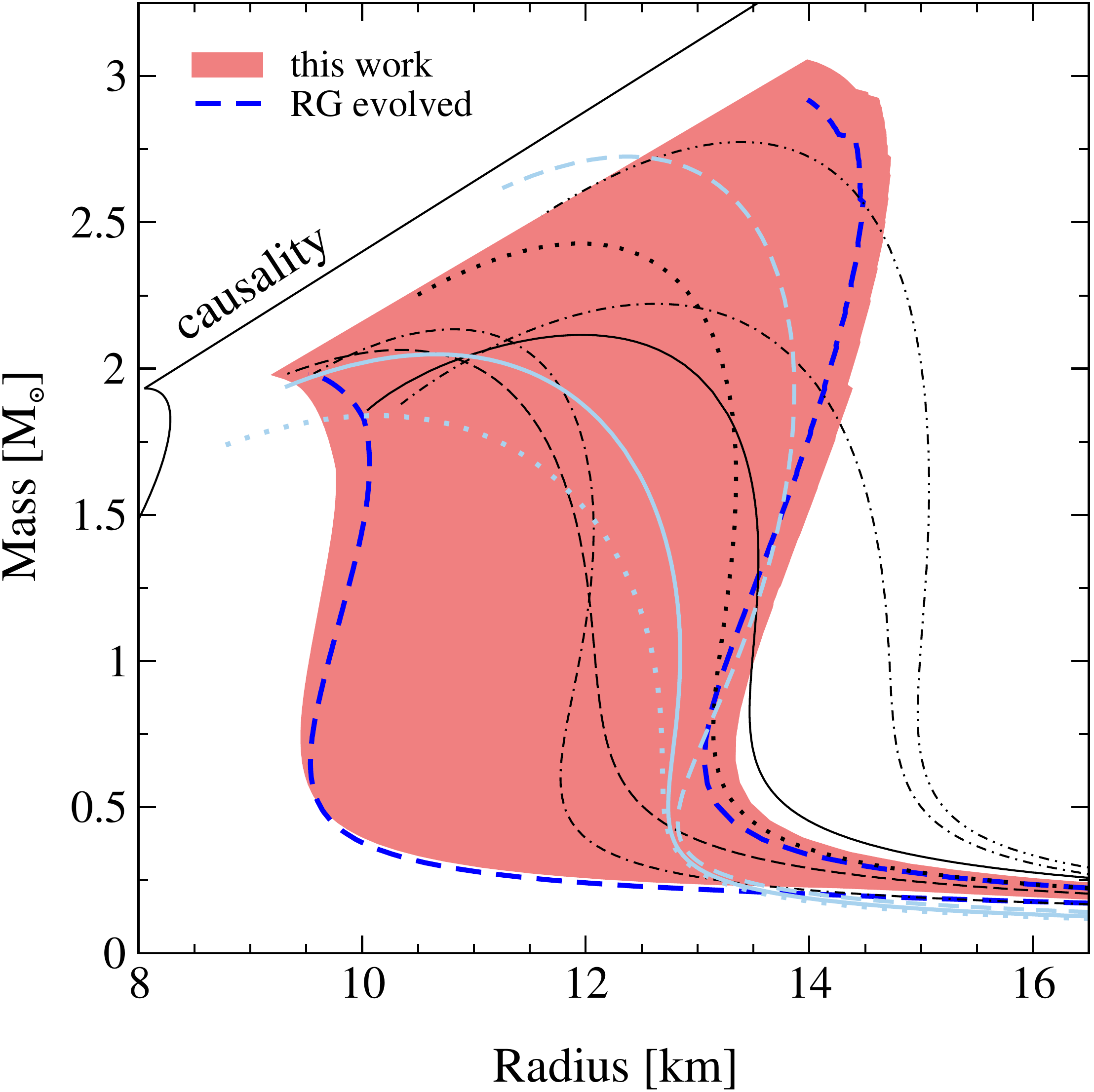}
\caption{Uncertainties in calculations of three nuclear systems. Left panel: Ground-state energies in the isotopic chains of Mg and Cl at second and third order in MBPT. The bands at each order are computed for five different chiral Hamiltonians~\cite{Simonis:2015vja}. Middle panel: Uncertainty band for the energy per particle of symmetric nuclear matter using MBPT. The colored and dashed bands illustrate the many-body uncertainty, while the individual bands are spanned by six chiral Hamiltonians~\cite{Drischler:2015eba}. Right panel: Uncertainty band for the mass-radius relation of neutron stars based on MBPT calculations of pure neutron matter and a polytropic high-density extension~\cite{Kruger:2013kua}.}
\label{fig:uncertainties}
\end{figure}

In the last years, there has been considerable progress in the description of nuclear systems due to the development of precision nuclear interactions and advances in ab-initio methods to solve the nuclear many-body problem. However, the predictions for nuclear observables are still limited by our insufficient understanding of the underlying nuclear forces and by our ability to reliably calculate strongly correlated systems, i.e., by approximations in the employed  many-body methods. In \cref{fig:uncertainties}, we show three examples of uncertainties in calculations of nuclear structure~\cite{Simonis:2015vja}, symmetric nuclear matter~\cite{Drischler:2015eba}, and the mass-radius relation of neutron stars~\cite{Kruger:2013kua}. The three calculations are based on many-body perturbation theory (MBPT) with chiral effective field theory (EFT) interactions. As it can be seen from the first two panels, current many-body uncertainties are much smaller than the uncertainty stemming from the nuclear Hamiltonian.

To explore the means of improving current uncertainties, it is desirable to have a consistent approach to nuclear systems ranging from nuclei to nucleonic matter, i.e., an approach that uses the same advanced many-body methods and the same systematic nuclear Hamiltonians and can provide reliable theoretical uncertainties. In this contribution, we discuss quantum Monte Carlo (QMC) results for nuclei up to $A=16$ and neutron matter, obtained by employing recently developed local chiral interactions. These interactions include consistent two- and three-nucleon forces up to next-to-next-to-leading-order (N$^2$LO), and have been fit to NN scattering and few-body observables probing the physics of light nuclei, with particular attention to $T=3/2$ physics. In addition, these interactions allow to estimate systematic uncertainties for nuclear systems. Our results show that such local chiral interactions give a very good description of the ground-state properties of nuclei (at least) up to \isotope[16]{O}, while providing an EOS of pure neutron matter compatible with astrophysical observations of neutron stars. This is an important step toward a predictive understanding of nuclei and neutron-star properties grounded in high-quality nuclear forces and ab-initio theory.

Our contribution is structured as follows. In \cref{sec:methods} we briefly review QMC methods and the employed local chiral interactions. We present results for nuclei and neutron matter in \cref{sec:results}. In \cref{sec:Issues} we address issues with local chiral interactions and discuss possible improvements. We summarize our work in  \cref{sec:summary}.

\section{Method and Hamiltonian}\label{sec:methods}

\subsection{Quantum Monte Carlo methods}\label{sec:QMC}

The solution of the many-body Schr\"odinger equation describing a system of interacting baryons is challenging because of the nonperturbative nature and the strong spin/isospin-dependence of realistic nuclear interactions. QMC methods provide a powerful tool to tackle the nuclear many-body problem in a nonperturbative fashion. They have proven to be remarkably successful in describing the properties of strongly correlated fermions in a large variety of physical conditions~\cite{Carlson:2015}.

In this contribution we present results obtained using the auxiliary field diffusion Monte Carlo (AFDMC) method~\cite{Schmidt:1999}, a stochastic technique developed to solve the many-body ground state of strongly correlated systems, such as nuclei and nuclear matter. The main idea is to evolve a many-body wave function in imaginary-time:
\begin{align}
	\Psi(\tau)=e^{-H\tau}\,\Psi_V \,,
\end{align}
where $H$ is the Hamiltonian of the system and $\Psi_V$ is a variational state
\begin{align}
    |\Psi_V\rangle=\big[F_C+F_2+F_3\big]|\Phi\rangle \,,
    \label{eq:psi}
\end{align}
where $F_C$ accounts for all the spin/isospin-independent correlations, and $F_2$ and $F_3$ are linear spin/isospin two- and three-body correlations as described in Ref.~\cite{Carlson:2015}. For nuclei, the term $|\Phi\rangle$ is taken to be a shell-model-like state with total angular momentum $J$, total isospin $T$, and parity $\pi$. Its wave function consists of a sum of Slater determinants $D$ constructed using single-particle orbitals:
\begin{align} 
    \langle RS |\Phi\rangle_{(J^\pi,T)} = \sum_n c_n\Big(\sum D\big\{\phi_\alpha(\vb{r}_i,s_i)\big\}\Big)_{(J^\pi,T)} \,,
\end{align}
where $\vb{r}_i$ are the spatial coordinates of the nucleons and $s_i$ represent their spins. Each single particle orbital $\phi_\alpha$ consists of a radial function $\varphi(r)$ coupled to the spin and isospin states. The determinants are coupled with Clebsch-Gordan coefficients to total $J$ and $T$, and the $c_n$ are variational parameters multiplying different components having the same quantum numbers. The radial functions $\varphi(r)$ are obtained by solving for the eigenfunctions of a Wood-Saxon well. For infinite matter, the term $|\Phi\rangle$ is built from a Slater determinant of plane waves with momenta $k_i$ quantized in a finite box whose volume is determined by the chosen density and number of particles involved. The infinite system is then realized by applying periodic boundary conditions~\cite{Gandolfi:2015jma}. All the parameters of $\Psi_V$ are chosen by minimizing the variational energy as described in Ref.~\cite{Sorella:2001}. 

After the optimization of $\Psi_V$, the trial wave function is propagated in imaginary time in order to remove all the low excited states that are still present in the variational ansatz, projecting the system onto the true ground state in the limit of $\tau\rightarrow\infty$. The evolution in imaginary-time is performed by sampling configurations of the system using Monte Carlo techniques. In more detail, in the AFDMC method both spatial coordinates and spin/isosospin configurations are sampled, the latter via a Hubbard-Stratonovich transformation. This is at variance with other imaginary time projection algorithms, such as the Green's function Monte Carlo (GFMC) method, where only the spatial degrees of freedom are sampled, and the propagation is carried out over all the possible good spin/isospin states of the system. The sign problem, that affects most of QMC algorithms for nuclear physics, is initially suppressed by evolving the wave function in imaginary time using the constrained-path approximation~\cite{Zhang:2003}. An unconstrained evolution is then performed until the sign-problem dominates and the variance of the results becomes severely large. Finally, expectation values are evaluated over the sampled configurations to compute the relevant observables. For more details see, for instance, Refs.~\cite{Carlson:2015,Lonardoni:2018prl,Lonardoni:2018prc,Lonardoni:2018nofk,Lynn:2019}.

In QMC calculations, nuclear systems are typically described as a collection of point-like particles of mass $m_N$ interacting via two- and three-body forces according to the nonrelativistic Hamiltonian
\begin{align}
	H=-\frac{\hbar^2}{2m_N}\sum_i \nabla_i^2+\sum_{i<j}v_{ij}+\sum_{i<j<k}V_{ijk} ,
\end{align}
where, in nuclei, the two-body interaction $v_{ij}$ also includes the Coulomb force. Historically, QMC methods have made use of phenomenological nuclear interactions, such as the Argonne $v_{18}$ (AV18) NN potential combined with Urbana/Illinois models for the three-nucleon (3N) forces~\cite{Carlson:2015}. By construction, these potentials are nearly local, meaning that the dominant parts of the interaction depend only on the relative distance $r$, spin, and isospin of the interacting nucleons, and not upon any derivatives. This feature is crucial for the application of QMC algorithms to the study of nuclear systems. In the last few years, with the development of local chiral interactions, a larger number of Hamiltonians has become accessible to QMC methods, providing new opportunities for nuclear structure and neutron-matter calculations.

\subsection{Local chiral interactions}\label{sec:LocChi}

Chiral NN interactions are typically derived in momentum space and depend on two momentum scales: the average momentum of the incoming particles $\vb{p}$ and the average momentum of the outgoing particles $\vb{p}'$. These momentum scales can be rewritten in terms of the momentum transfer $\vb{q}=\vb{p}'-\vb{p}$ and the momentum transfer in the exchange channel, $\vb{k}=(\vb{p}+\vb{p}')/2$. Upon Fourier transformation, all dependencies on $\vb{q}$ transform to dependencies on the interparticle distance $r$, while dependencies on $\vb{k}$ lead to derivatives and nonlocalities. 

The typical sources of nonlocalities are: (i) common regulator functions of the form $f(p)=\exp(-(p/\Lambda)^{2n})$ for both $p$ and $p'$, where $\Lambda$ is the cutoff scale and $n$ is an integer number, and (ii) $\vb{k}$-dependent operators in the potentials. In general, chiral interactions are written in terms of pion-exchange and contact contributions,
\begin{align}
    V(\vb{q},\vb{k})=V_{\rm{cont}}(\vb{q},\vb{k})\cdot f_{\rm{short}}(\vb{q},\vb{k})+V_{\pi}(\vb{q},\vb{k})\cdot f_{\rm{long}}(\vb{q},\vb{k})\,,
\end{align}
where the functions $f_\alpha$ denote the short- and long-range regulator functions. In order to remove the first source of nonlocality, we construct chiral interactions with local regulators $f_\alpha(\vb{q})$. In particular, we use the $r$-dependent coordinate-space regulator functions
\begin{align}
    f_{\rm{long}}(r)&=\left(1- e^{ -\left(\frac{r}{R_0} \right)^{n_1}} \right)^{n_2} \,,\\
    f_{\rm{short}}(r)&=\frac{n}{4 \pi\,R_0^3\,\Gamma\left(\frac{3}{n}\right)}\,e^{-\left(\frac{r}{R_0} \right)^{n}} \,.
\end{align} 

For the second source of nonlocalities, we first stress that pion-exchange interactions are local up to N$^2$LO: $V_{\pi}^{\rm N^2LO}(\vb{q},\vb{k})\equiv V_{\pi}^{\rm N^2LO}(\vb{q})$. In coordinate space, these interactions can be written in the form
\begin{align}
    V_{\pi}(r)=V_C(r)+W_C(r)\,\bm\tau_1\cdot\bm\tau_2 +\big(V_S(r)+W_S(r)\,\bm\tau_1\cdot\bm\tau_2\big)\,\bm\sigma_1\cdot\bm\sigma_2 +\big(V_T(r)+W_T(r)\,\bm\tau_1\cdot\bm\tau_2\big)\,S_{12}\,
\end{align}
where $S_{12}=3\,\bm\sigma_1\cdot\hat{\vb{r}}\;\bm\sigma_2\cdot\hat{\vb{r}}-\bm\sigma_1\cdot\bm\sigma_2$ is the tensor operator. Applying a local regulator to $V_\pi(r)$ keeps its contributions local, and the only remaining source of nonlocalities is $V_{\rm{cont}}(\vb{q},\vb{k})$. To remove those, we make use of ambiguities in the general operator basis of the contact operators at a certain chiral order~\cite{Gezerlis:2013,Gezerlis:2014}. For example, at leading order (LO) in the chiral expansion, the most general operator basis is given by
\begin{align}
    V_{\rm cont}^{\rm LO}(\vb{q},\vb{k})=V_{\rm cont}^{\rm LO}=\alpha_1 \mathbbm{1}+ \alpha_2\,\bm\sigma_1\cdot\bm\sigma_2+\alpha_3\,\bm\tau_1\cdot\bm\tau_2+ \alpha_4\,\bm\sigma_1\cdot\bm\sigma_2\,\bm\tau_1\cdot\bm\tau_2\,,
    \label{eq:LOpot}
\end{align}
where the $\alpha_i$ are the LO low-energy couplings (LECs). However, due to the Pauli principle, only two out these four operators are linearly independent, as it can be seen by applying the antisymmetrization operator to the potential; see also Sec.~\ref{sec:Fierz}. Therefore, one can choose any pair out of these four operators for the LO contact potential, which leads to ambiguities in the contact interactions similar to Fierz ambiguities~\cite{Huth:2017}. A common choice is given by the first two operators, 
\begin{equation}
V_{\rm cont}^{\rm LO}=C_S \mathbbm{1}+ C_T\,\bm\sigma_1\cdot\bm\sigma_2 \,.
\end{equation}

At next-to-leading order (NLO), the contact interaction is momentum-dependent, and it is given by
\begin{align}
    V_ {\rm cont}^{\rm NLO}(\vb{q},\vb{k}) = & \gamma_1 \, q^2 + \gamma_2 \, q^2\, {\bm \sigma}_1 \cdot {\bm \sigma}_2 + \gamma_3 \, q^2\, {\bm \tau}_1 \cdot {\bm \tau}_2  + \gamma_4 \, q^2 {\bm \sigma}_1 \cdot {\bm \sigma}_2 {\bm \tau}_1 \cdot {\bm \tau}_2 + \gamma_5 \, k^2 + \gamma_6 \, k^2\, {\bm \sigma}_1 \cdot {\bm \sigma}_2 \nonumber \\
    + & \gamma_7 \, k^2\, {\bm \tau}_1 \cdot {\bm \tau}_2 + \gamma_8 \, k^2 {\bm \sigma}_1 \cdot {\bm \sigma}_2 {\bm \tau}_1 \cdot {\bm \tau}_2 + \gamma_9 \, ({\bm \sigma}_1 + {\bm \sigma}_2)({\bf q}\times {\bf k}) + \gamma_{10} \, ({\bm \sigma}_1 + {\bm \sigma}_2)({\bf q}\times {\bf k}) {\bm \tau}_1 \cdot {\bm \tau}_2 \nonumber \\
    + & \gamma_{11} ({\bm \sigma}_1 \cdot {\bf q}) ({\bm \sigma}_2 \cdot {\bf q}) + \gamma_{12} ({\bm \sigma}_1 \cdot {\bf q}) ({\bm \sigma}_2 \cdot {\bf q}) {\bm \tau}_1 \cdot {\bm \tau}_2  + \gamma_{13} ({\bm \sigma}_1 \cdot {\bf k}) ({\bm \sigma}_2 \cdot {\bf k}) \nonumber \\
    + & \gamma_{14} ({\bm \sigma}_1 \cdot {\bf k}) ({\bm \sigma}_2 \cdot {\bf k}) {\bm \tau}_1 \cdot {\bm \tau}_2 \,,
\end{align}
where the $\gamma_{i}$ are again a set of LECs. Using the same arguments as before, we choose all local contact interactions $\propto q^2$, as well as the nonlocal spin-orbit interaction, that can, however, be treated within QMC methods. 

At N$^2$LO, the short-range operator structure of the potential is the same as that at NLO, with no additional momentum dependences. The result is then a local chiral interaction up to N$^2$LO that can be efficiently used in QMC methods. The appearing LECs, that accompany all of the short-range operators, have been fit to reproduce NN scattering phase shifts, and we refer to Ref.~\cite{Gezerlis:2014} for details. 

\begin{figure}[htb]
\begin{minipage}[h]{0.5\textwidth}
\centering
\includegraphics[width=0.95\textwidth]{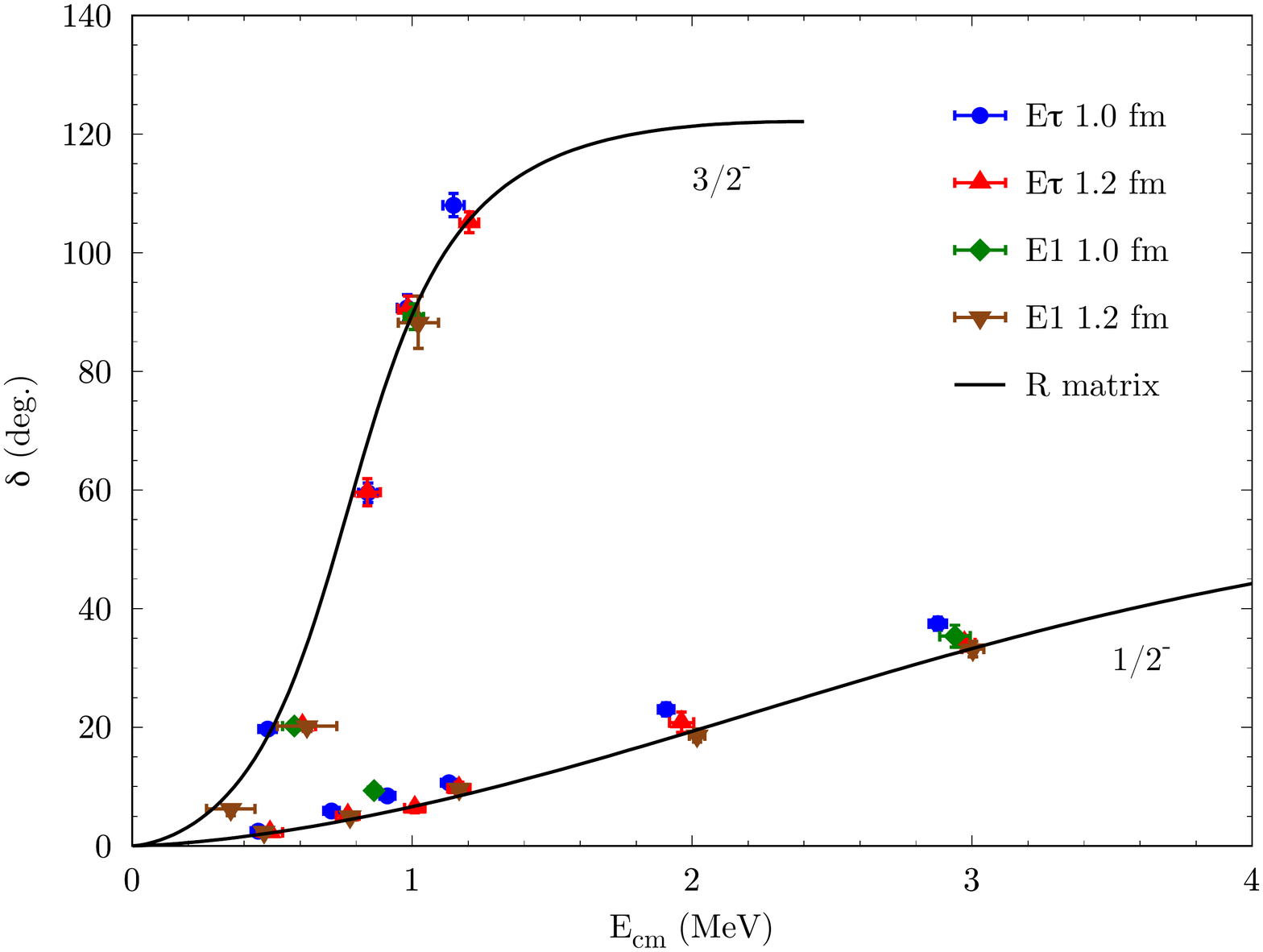}
\end{minipage}
\begin{minipage}[h]{0.5\textwidth}
\centering
\includegraphics[width=0.65\textwidth]{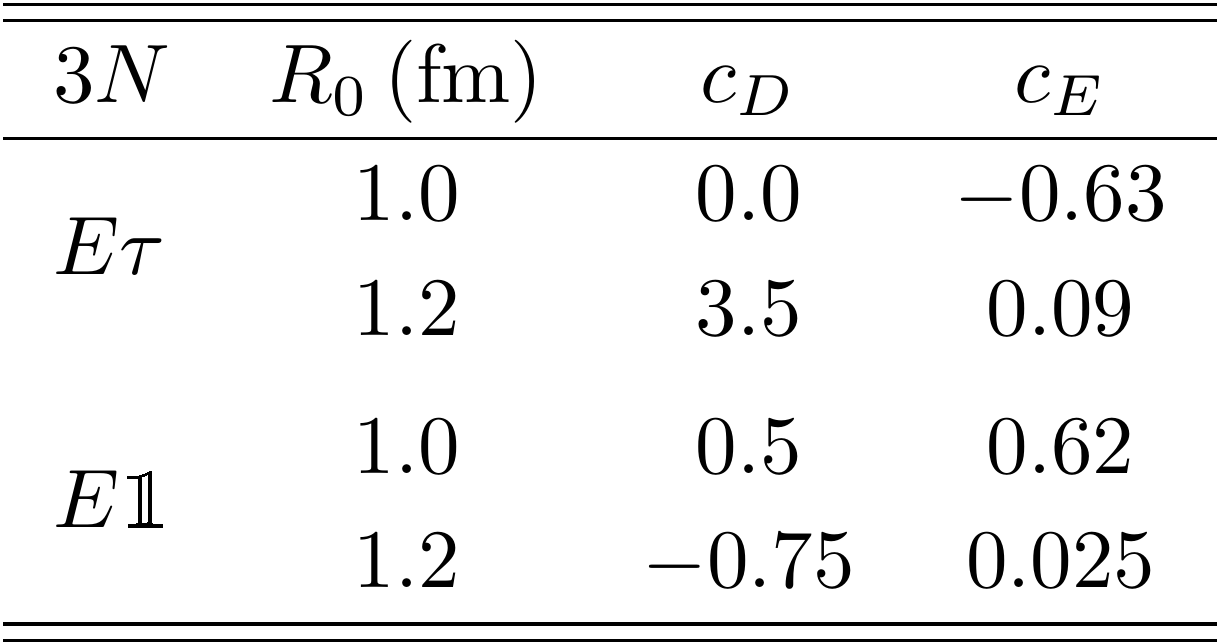}
\end{minipage}
\caption[]{Left panel: $P$-wave $n$-$\alpha$ elastic scattering phase shifts for local chiral interactions at N$^2$LO~\cite{Lonardoni:2018prc} compared to an $R$-matrix analysis of experimental data~\cite{Hale}. Right panel: LECs $c_D$ and $c_E$ for different coordinate-space cutoffs and parametrizations of the 3N contact term $V_E$~\cite{Lonardoni:2018prc}.}
\label{fig:nalpha}
\end{figure}

In addition to NN interactions, 3N forces also naturally appear at N$^2$LO. These can be grouped into three topologies: (i) a 3N two-pion exchange (TPE) interaction, labeled $V_C$, which depends on the NN LECs $c_i$; (ii) a one-pion-exchange--contact interaction, labeled $V_D$, which depends on a true 3N coupling $c_D$; (iii) a 3N contact interaction $V_E$, which depends on a second true 3N LEC $c_E$. The TPE interaction consists of an $S$-wave and $P$-wave contributions, where the latter is the well-known Fujita-Miyazawa interaction. For the 3N contact term $V_E$, the most general operator basis consists of six different operators which reduce to only one linearly-independent term upon antisymmetrization; see Sec.~\ref{sec:Fierz}. However, regulator artifacts, which we discuss later, lead to ambiguities. We will study the impact of these regulator artifacts by investigating three different operator choices, namely $E\tau$, $E\mathbbm1$, and $E P$; see Sec.~\ref{sec:Fierz} and Refs.~\cite{Lynn:2016,Lynn:2017fxg,Lonardoni:2018prc} for more details. The leading 3N interactions at N$^2$LO are local, if local regulator functions with 3N cutoff $R_{3N}=R_0$ are employed, and they can be easily implemented in QMC methods~\cite{Tews:2016}. For the local chiral interactions used in this work, the unknown 3N LECs $c_D$ and $c_E$ have been fit to reproduce the binding energy of \isotope[4]{He} as well as the $P$-wave $n$-$\alpha$ elastic scattering phase shifts using the GFMC method~\cite{Lynn:2016,Lonardoni:2018prc}, see \cref{fig:nalpha}. 

\section{Results}\label{sec:results}
Recent advances made in accurate QMC methods and their combination with interactions derived from chiral EFT have provided many new insights in low-energy nuclear theory~\cite{Lynn:2019}. A remarkable result is the possibility to describe nuclei with $A\leq 16$ and their global properties from microscopic nuclear Hamiltonians constructed to reproduce only few-body observables, while simultaneously predicting properties of neutron stars compatible with astrophysical observations~\cite{Lynn:2016}. In the next two sections, we summarize the nuclear-structure results for light- and medium-mass nuclei, and the predictions for neutron matter.

\subsection{Nuclei}

\begin{figure}[b]
\centering
\includegraphics[width=0.495\textwidth]{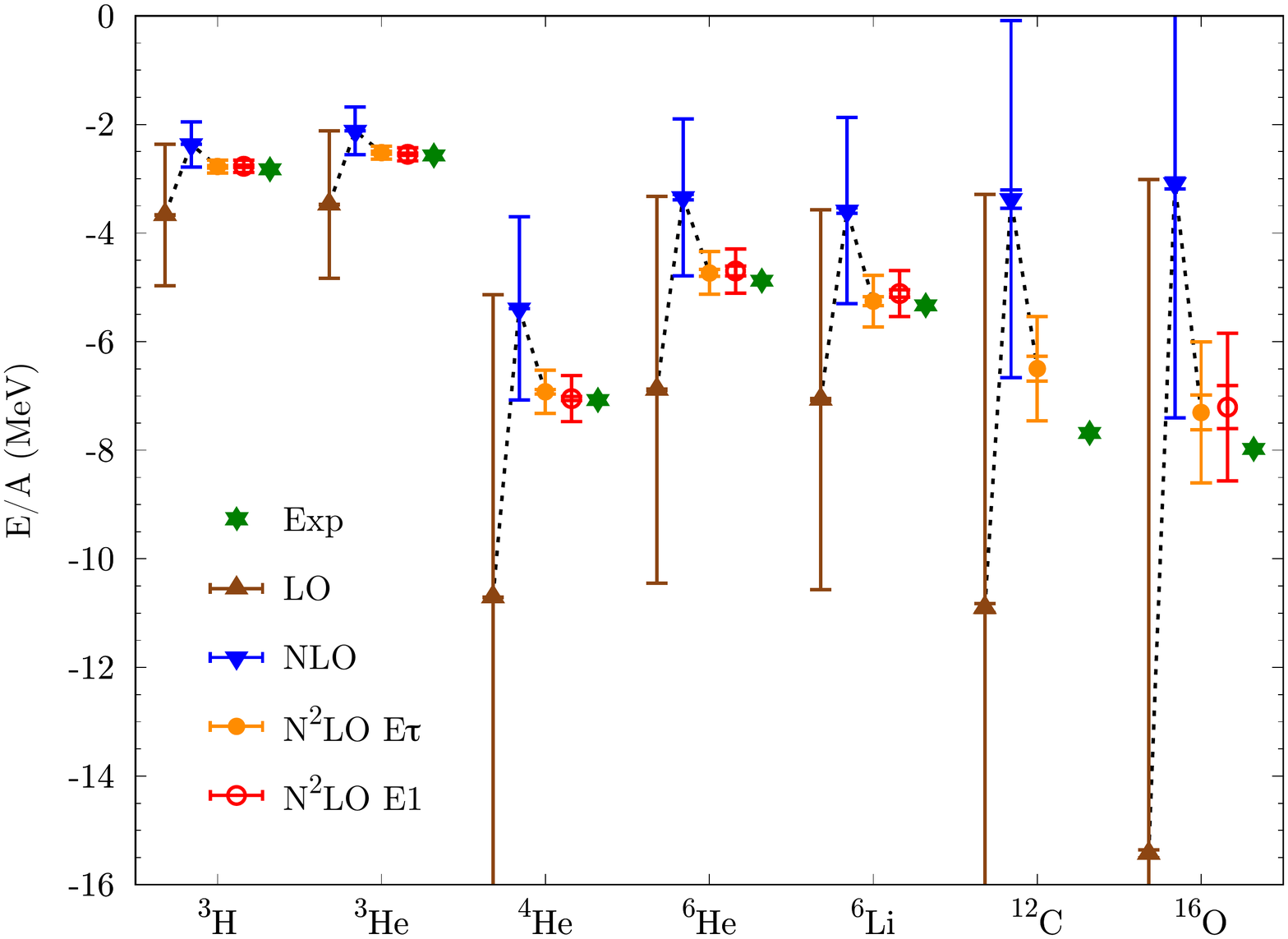}
\includegraphics[width=0.495\textwidth]{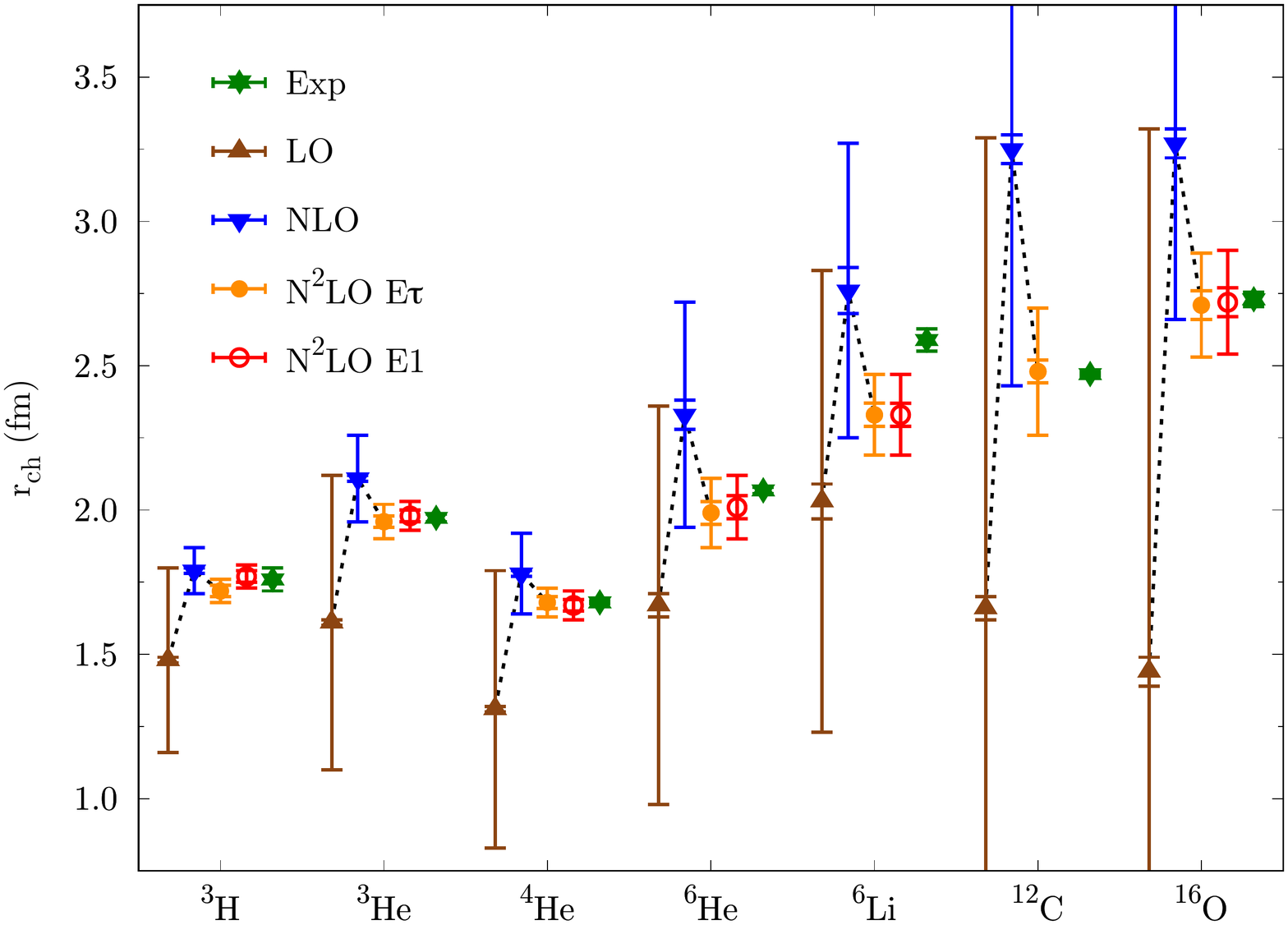}
\caption[]{Ground state properties for nuclei with $3\le A\le16$ using local chiral potentials~\cite{Lonardoni:2018prl,Lonardoni:2018prc}. Left panel: Binding energy per nucleon. Right panel: Charge radius. Results at different order of the chiral expansion and for different parametrizations of the 3N contact term $V_E$ are shown for the coordinate-space cutoff $R_0=1.0\,\rm fm$. Smaller error bars indicate the statistical Monte Carlo uncertainty, while larger error bars are the uncertainties coming from the truncation of the chiral expansion.}
\label{fig:ene_rch}
\end{figure}

We have employed local chiral potentials up to N$^2$LO in AFDMC calculations of ground-state properties of nuclei up to \isotope[16]{O}~\cite{Lonardoni:2018prl,Lonardoni:2018prc}. In~\cref{fig:ene_rch} we show the results for the binding energy (left panel) and the charge radius (right panel) for different systems for the coordinate-space cutoff $R_0=1.0\,\rm fm$ (harder interaction). AFDMC predictions are shown for each order of the chiral expansion (upward brown triangle for LO, downward blue triangle for NLO, orange and red circles for N$^2$LO), compared to experimental results (green stars). At N$^2$LO, results are obtained for different operator structures of the 3N contact term $V_E$, namely $E\tau$ (solid orange circles) and $E\mathbbm1$ (empty red circles). Smaller error bars indicate the statistical Monte Carlo uncertainty, while larger error bars are the uncertainties coming from the truncation of the chiral expansion according to the EKM prescription~\cite{Epelbaum:2015epja}. \cref{fig:ene_rch} summarizes several original contributions of this study: (i) AFDMC is used for the first time to calculate properties of closed- and open-shell nuclei up to $A=16$ using realistic two- and three-body potentials; (ii) a complete quantification of all uncertainties associated to the employed Hamiltonian and nuclear many-body method is provided; (iii) the computed observables manifest a good order-by-order convergence pattern in the chiral expansion; (iv) both energies and radii at N$^2$LO are well reproduced up to $A=16$, even though only few-body physics has been used to fit the employed potentials; (v) different three-body operator structures at N$^2$LO provide the same description of the analyzed observables, i.e., regulator artifacts coming from the violation of the Fierz-rearrangement freedom in the selection of local contact operators (cf. Sec.~\ref{sec:Fierz}) are small for this cutoff scale. Similar conclusions are found for the cutoff $R_0=1.2\,\rm fm$ (softer interaction) up to $A=6$. Peculiar is the case of \isotope[16]{O} for such a softer potential, for which the nucleus is significantly overbound and very compact. In addition, regulator artifacts are much larger in this case. See Refs.~\cite{Lonardoni:2018prl,Lonardoni:2018prc} for a complete discussion.

\begin{figure}[t]
\centering
\includegraphics[width=0.495\textwidth]{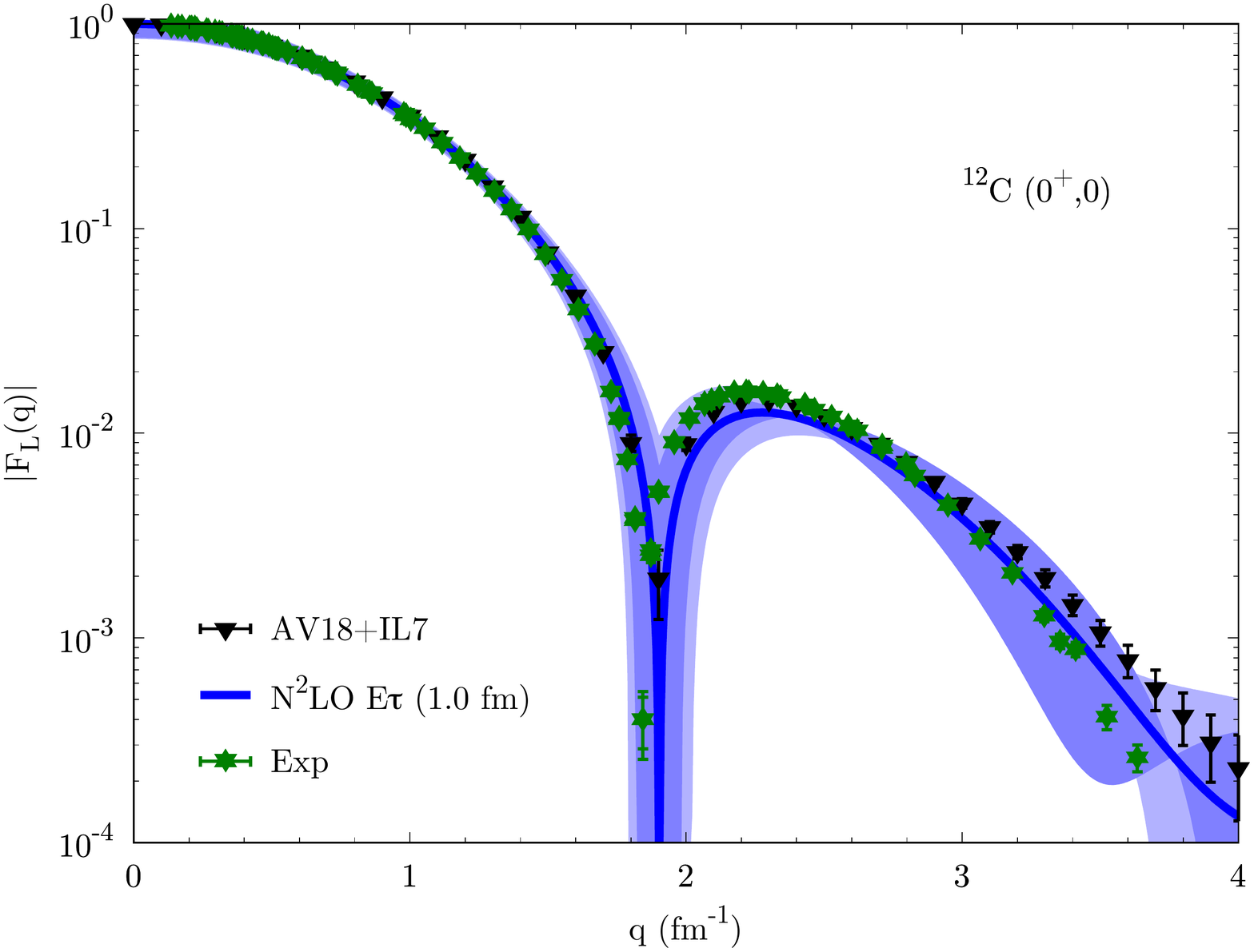}
\includegraphics[width=0.495\textwidth]{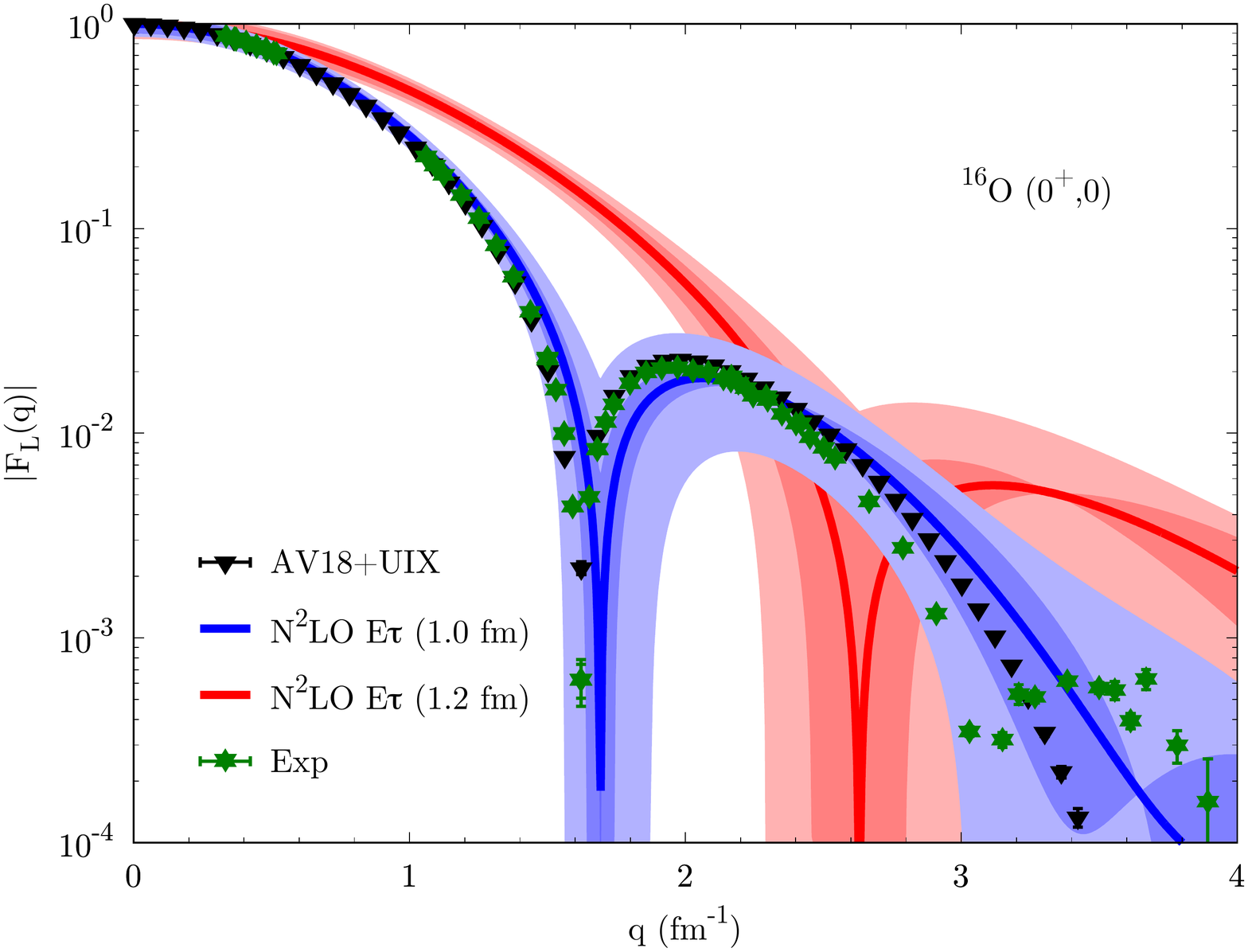}
\caption[]{Charge form factors using local chiral interactions at N$^2$LO for different parametrizations of the 3N contact term $V_E$ and different coordinate-space cutoffs~\cite{Lonardoni:2018prl,Lonardoni:2018prc}. Colored bands correspond to uncertainties coming from the truncation of the chiral expansion. Black triangles are results employing phenomenological potentials~\cite{Lovato:2013,Lonardoni:2017}. Experimental data are based on Refs.~\cite{Devries:1987,Sick:1970,Schuetz:1975,Sick:1975}. Left panel: \isotope[12]{C}. Right panel:~\isotope[16]{O}.}
\label{fig:ff}
\end{figure}

In~\cref{fig:ff} we present the results for the charge form factor in \isotope[12]{C} (left panel) and \isotope[16]{O} (right panel). Colored lines and bands refer to results employing local chiral interactions at N$^2$LO, black triangles are QMC results for phenomenological potentials, and green stars are the experimental data. Within the estimated uncertainties (both statistical and systematic), the harder local chiral potential $(R_0=1.0\,\rm fm)$ provides an excellent description of the charge form factor in both systems. Different is the case of the softer interaction $(R_0=1.2\,\rm fm)$ in \isotope[16]{O}, for which the first diffraction minimum occurs at a significantly higher momentum than experimentally observed. This is, however, consistent with the \isotope[16]{O} overbinding and compactness obtained for this interaction (see Refs.~\cite{Lonardoni:2018prl,Lonardoni:2018prc} for more details).

\begin{figure}[t]
\centering
\includegraphics[width=0.495\textwidth]{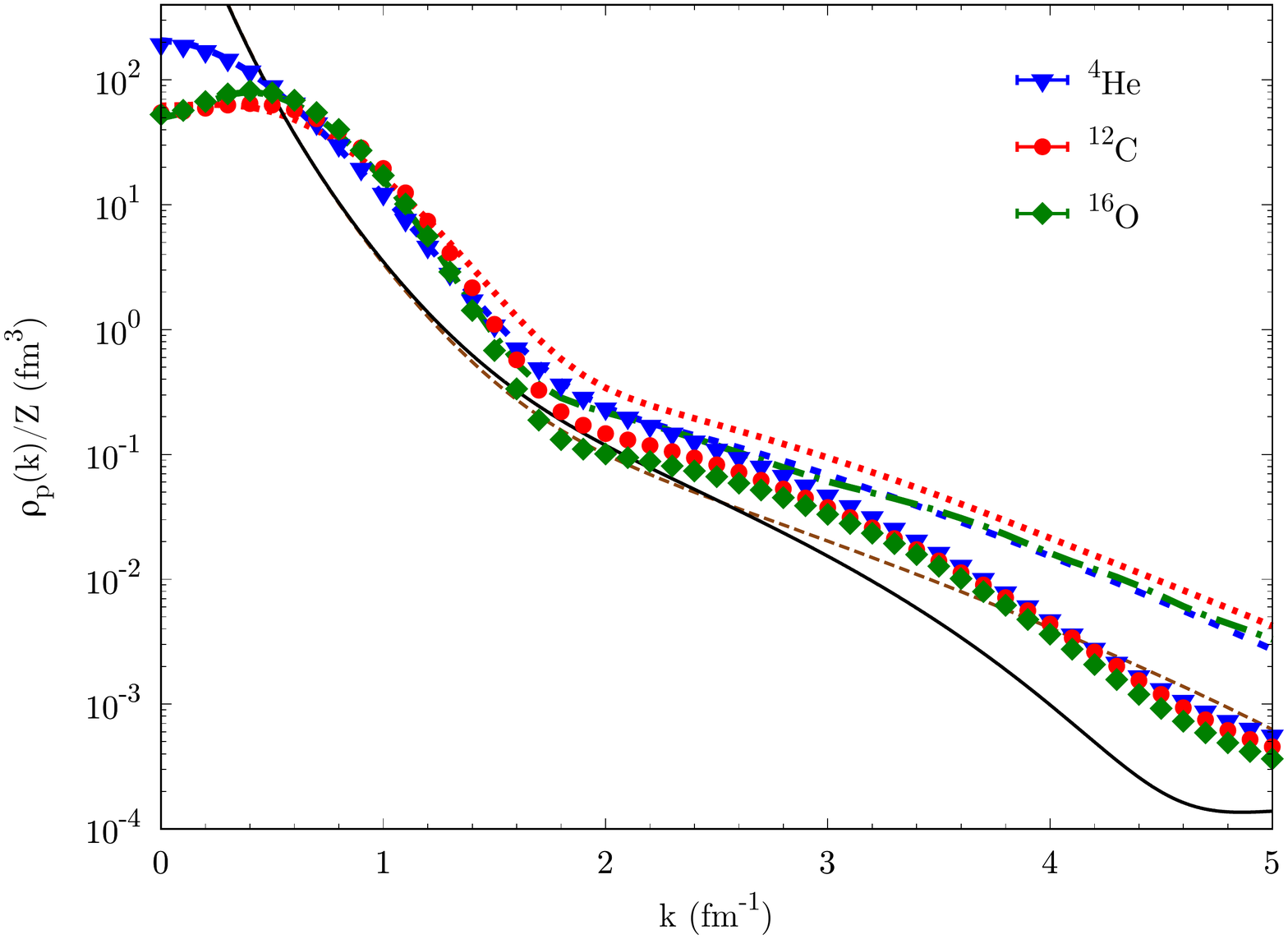}
\includegraphics[width=0.495\textwidth]{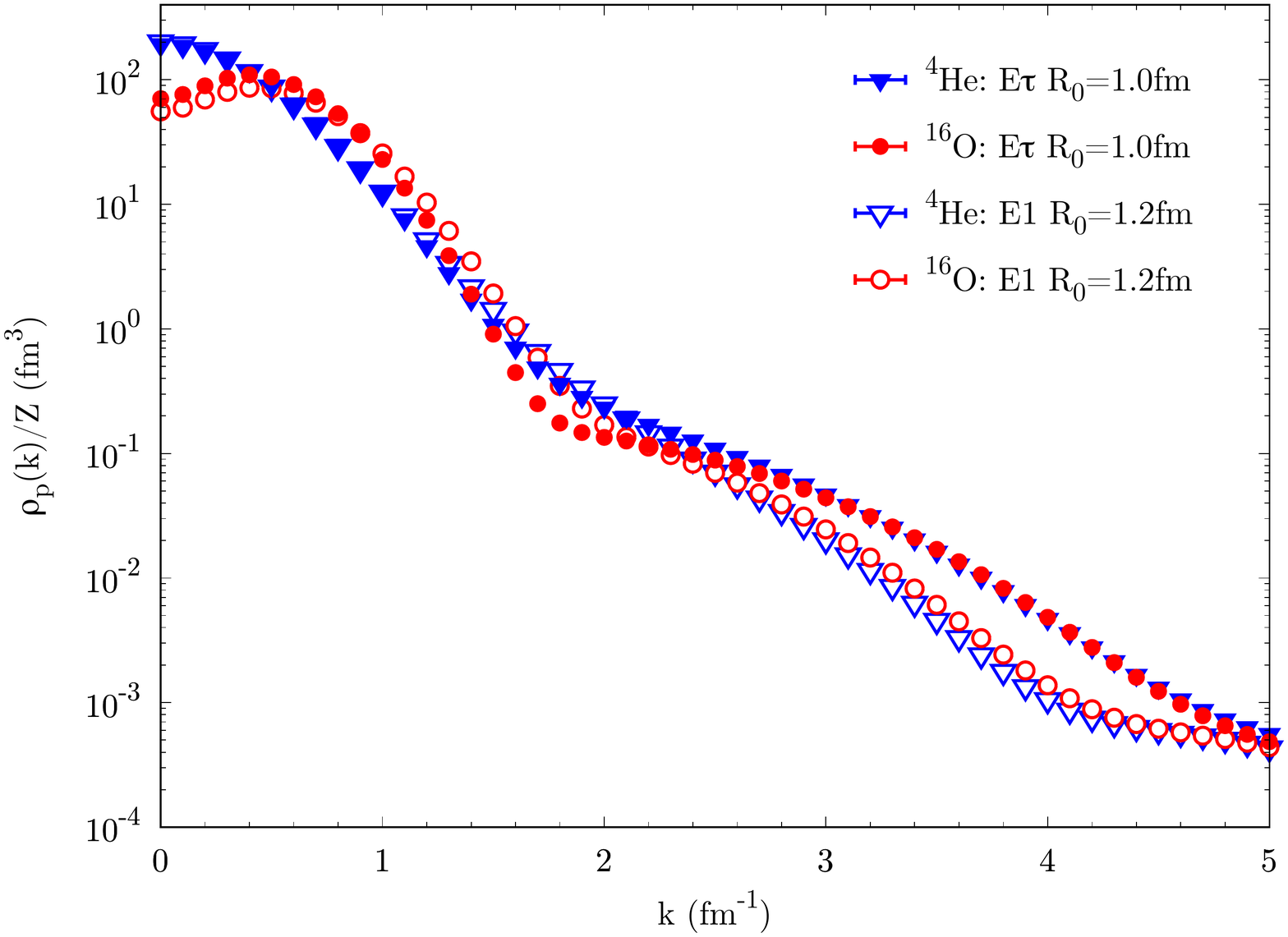}
\caption[]{Proton momentum distributions for local chiral interactions at N$^2$LO~\cite{Lonardoni:2018nofk}. Left panel: \isotope[4]{He}, \isotope[12]{C}, and \isotope[16]{O} (solid symbols) compared to phenomenological potentials (dashed lines). The deuteron momentum distributions are also shown. Right panel: \isotope[4]{He} and \isotope[16]{O}, solid (empty) symbols for the $E\tau$ ($E\mathbbm1$) parametrization of the 3N contact term with cutoff $R_0=1.0\,\rm fm$ ($R_0=1.2\,\rm fm$).}
\label{fig:nofk}
\end{figure}

QMC methods have also been used to study short range correlation (SRC) effects as emerging from the underlying microscopic Hamiltonian. We have performed a variational Monte Carlo study of single-nucleon momentum distributions in $A\le16$ nuclei~\cite{Lonardoni:2018nofk}. \cref{fig:nofk} shows the results for different nuclei and for different interaction schemes. In the left panel, chiral results at N$^2$LO for coordinate-space cutoff $R_0=1.0\,\rm fm$ (solid symbols) are compared to results for phenomenological potentials (dashed lines) for $A=4,12,16$. In the right panel, results for \isotope[4]{He} (blue triangles) and \isotope[16]{O} (red circles) are compared for different local chiral interactions at N$^2$LO: solid symbols for $R_0=1.0\,\rm fm$, empty symbols for $R_0=1.2\,\rm fm$. We find that the single-nucleon momentum distribution manifests the expected universal behavior at high momentum, i.e., the independence of the high-momentum components upon the specific nucleus. However, such a universal behavior appears to be scheme dependent, i.e., it depends on the choice of the employed potential since it is determined by the short-range structure of the selected Hamiltonian. See Ref.~\cite{Lonardoni:2018nofk} for a complete discussion. The prospect of directly probing nucleon momentum distributions in light nuclei via $(e,e'p)$ measurements has also been explored from a theoretical point of view within our QMC framework~\cite{Cruz:2019}.

\begin{figure}[b]
\centering
\includegraphics[width=0.495\textwidth]{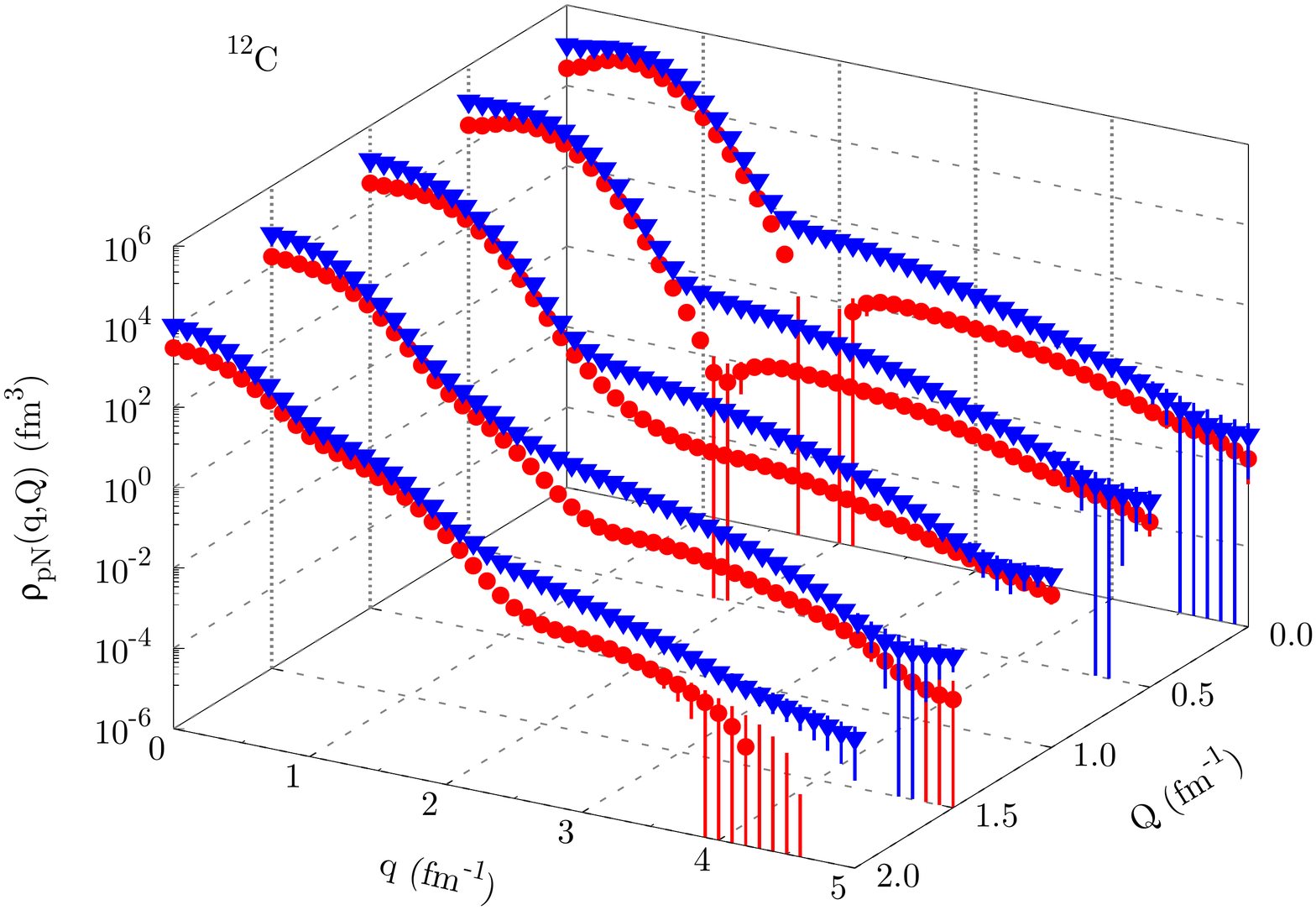}
\includegraphics[width=0.495\textwidth]{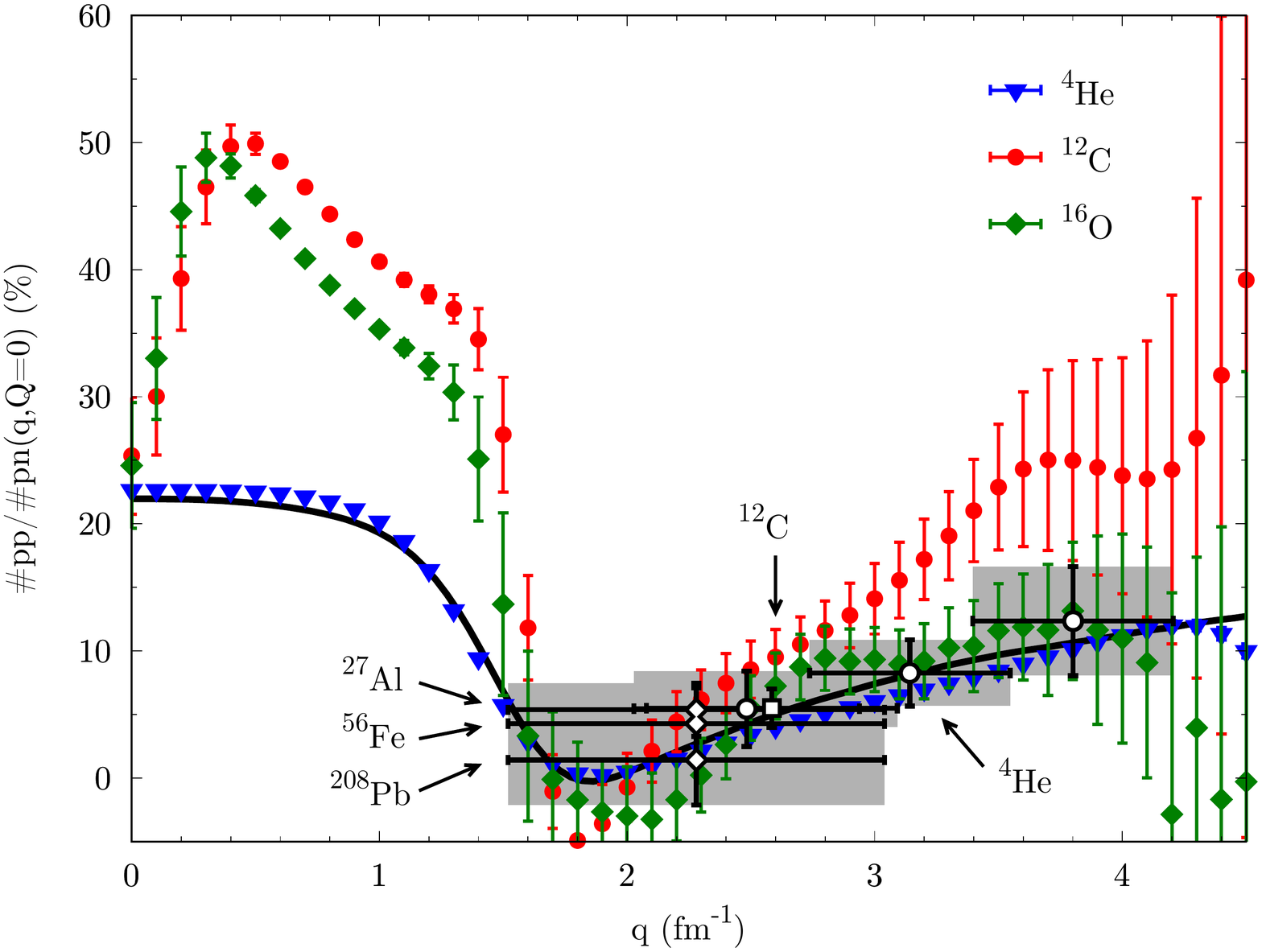}
\caption[]{Results extracted from the analysis of two-nucleon momentum distributions for the N$^2$LO $E\tau$ potential with cutoff $R_0=1.0\,\rm fm$. Left panel: Distributions in \isotope[12]{C}, blue triangles for $pn$ pairs, red circles for $pp$ pairs~\cite{Lonardoni:2018nofk}. Right panel: $pp$ to $pn$ pairs ratio in $A=4,12,16$ nuclei as a function of the relative momentum $q$ for back-to-back $(Q=0)$ pairs~\cite{Lonardoni:2018nofk}. Black symbols are extracted from experimental data~\cite{Subedi:2008,Korover:2014,Hen:2014}.}
\label{fig:nofq}
\end{figure}

In addition to single-nucleon momentum distribution, we have used variational Monte Carlo techniques to calculate two-nucleon momentum distributions in $A=4,12,16$ nuclei~\cite{Lonardoni:2018nofk}. In the left panel of~\cref{fig:nofq}, we present the proton-proton (red circles) and proton-neutron (blues triangles) momentum distributions in \isotope[12]{C} as a function of both the relative momentum $q$ of the pair and the center of mass momentum $Q$. Results are shown for the N$^2$LO $E\tau$ local chiral potential with cutoff $R_0=1.0\,\rm fm$. At $Q=0$ $pn$ pairs show a deuteronlike distribution, with a change of slope around $q=1.5\,\rm fm^{-1}$, as found for phenomenological potentials~\cite{Wiringa:2014}. The $pp$ distribution presents instead a node in this region, located at approximately $2\,\rm fm^{-1}$. The same qualitative conclusions hold for different systems and different interaction schemes. It follows that $\rho_{N\!N}(q,Q=0)$ is larger for $pn$ pairs compared to $pp$ pairs, in particular for relative momenta in the range $q\approx1.5-2.5\,\rm fm^{-1}$. This is consistent with the observation of large differences in the $pp$ and $pn$ distributions at moderate values of the relative momentum extracted from $(e,e'pN)$ experiments~\cite{Subedi:2008,Korover:2014,Hen:2014}. Moreover, even though momentum distributions themselves are model dependent, the ratio of $pp$ to $pn$ back-to-back pairs appears to be largely model independent and matches the available experimental information extracted from $(e,e'pN)$ experiments from light to heavy nuclei (see right panel of~\cref{fig:nofq}). 

Another interesting opportunity arising from the use of QMC methods with EFT techniques is the possibility of naturally explaining the empirical linear relationship between the slope of the EMC effect in deep inelastic scattering and the SRC scaling factors $a_2$ in quasi-elastic lepton-nucleus scattering, allowing us to calculate and predict SRC scaling factors from ab-initio low-energy nuclear theory~\cite{Chen:2017,Lynn:2019a2}. We have performed diffusion Monte Carlo calculations of nuclei from $A=2$ to $A=40$ using both phenomenological and local chiral interactions and deriving $a_2$ values from the ratio of two-nucleon densities in coordinate space. Results show that, even though two-nucleon densities are scheme and scale dependent quantities, their ratio, as predicted from EFT up to higher-order corrections, is scheme and scale independent and in good agreement with available experimental data~\cite{Lynn:2019a2}. 

\subsection{Neutron Matter}\label{sec:neutronmatter}

The AFDMC method has also been extensively used to study pure neutron systems, including neutron drops and pure neutron matter (PNM)~\cite{Gandolfi:2011,Gandolfi:2014_epja,Gandolfi:2015jma,Gandolfi:2017}. In this section, we focus on the results for PNM obtained by employing the same local chiral interactions described in the previous sections.

\begin{figure}[t]
\centering
\includegraphics[width=0.45\textwidth]{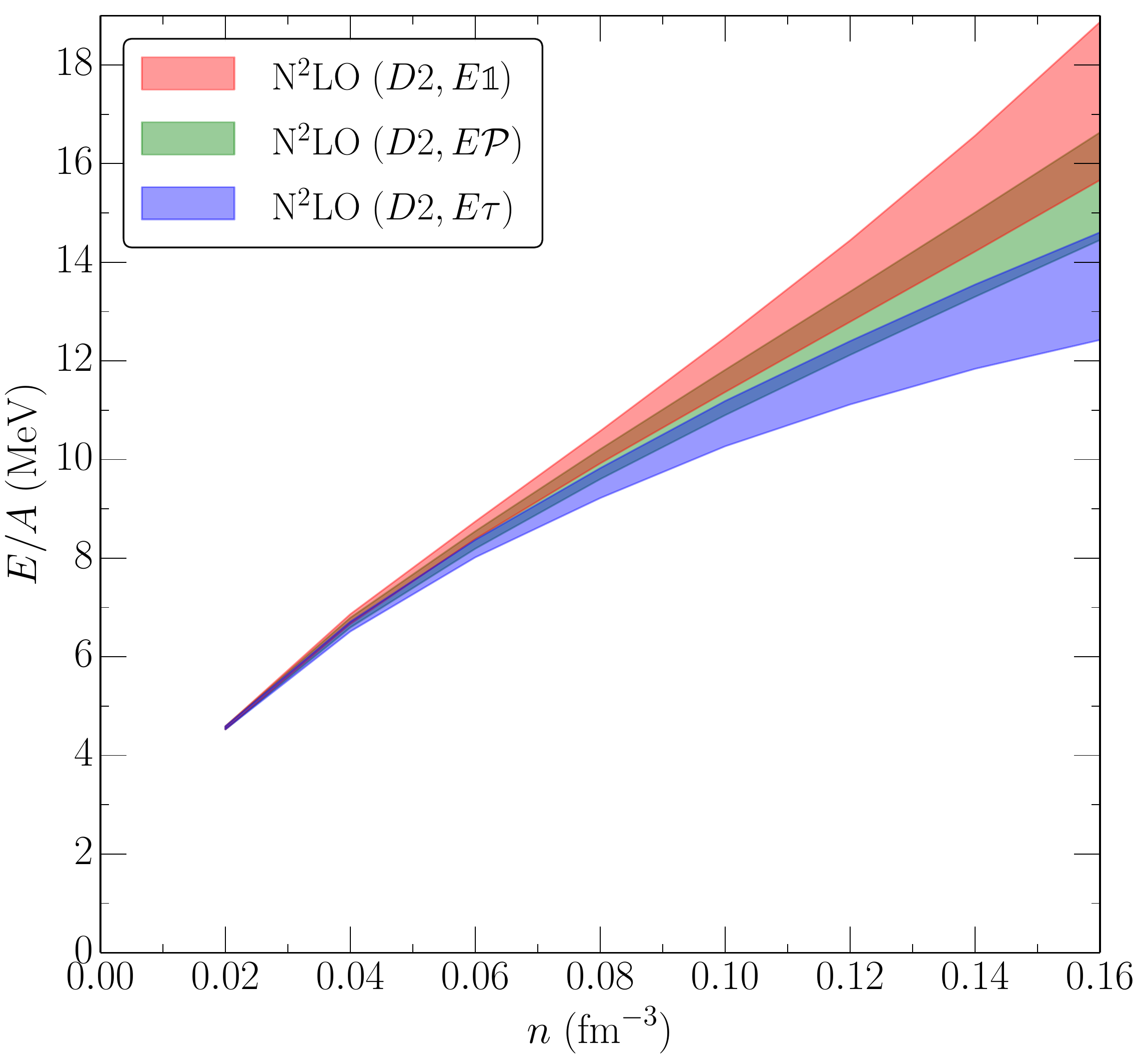}
\caption[]{EOS of PNM for different choices of the 3N contact operator in $V_E$ and cutoff $R_0=1.0\,\rm fm$~\cite{Lynn:2016}. The individual colored bands correspond to uncertainties stemming from the truncation of the chiral expansion.}
\label{fig:eos}
\end{figure}

In \cref{fig:eos}, we show the equation of state of PNM, i.e., the energy per particle as a function of the baryon density $n$, as obtained using local chiral interactions at N$^2$LO with coordinate-space cutoff $R_0=1.0\,\rm fm$. In particular, three uncertainty bands are shown. While each band is estimated according to the EKM prescription~\cite{Epelbaum:2015epja} using the average momentum in a Fermi gas, $k_{\text{avg}}=\sqrt{3/5}\,k_F$, as the characteristic momentum scale, the three different bands explore the uncertainty due to regulator artifacts stemming from the 3N contact interaction $V_E$. Typically, only the 3N TPE interaction $V_C$ contributes to PNM, as the shorter-range topologies $V_E$ and $V_D$ vanish in $T=3/2$ or $S=3/2$ systems due to the Pauli principle and their spin-structure~\cite{Hebeler:2009iv}. However, local regulators smear out these delta-like contact interactions over a finite volume. As a result, their contribution to the energy per particle of PNM is non zero; see Sec.~\ref{sec:Fierz} for a detailed discussion. 

The uncertainty due to regulator artifacts is $\approx 4\,\rm MeV/A$ at nuclear saturation density, $n_0=0.16\,\rm fm^{-3}$, which is rather sizable and larger than the truncation-uncertainty estimate of $\approx 3\,\rm MeV/A$. In total, we find an energy per particle in the range $\approx 12-19\,\rm MeV$ at $n_0$. We have explored the EOS of PNM at even higher densities in Ref.~\cite{Tews:2018kmu}, finding that the uncertainties increase fast as a function of the baryon density. At twice nuclear saturation density, the total uncertainty is $\approx 6-42\,\rm MeV$.

We have used the AFDMC neutron-matter calculations to study neutron-star--structure observables and the recent neutron-star merger in Refs.~\cite{Tews:2018kmu,Tews:2018chv,Tews:2019cap}. In these studies, we extend the AFDMC calculations of PNM to neutron-star conditions, i.e., we extend the results to $\beta$ equilibrium and include a neutron-star crust. We then use these constraints up to a density $n_{\rm tr}$, which we vary in the range of $1-2 n_0$, and extend the results to even higher densities using a speed-of-sound extension. We find that, even though uncertainties grow quite fast with density, EOS constraints in the density range from $1-2 n_0$ are very valuable to constrain neutron-star radii and the gravitational-wave signal from neutron-star mergers. Using the EOS up to saturation density as a constraint, we find the radius of a typical $1.4 M_{\odot}$ neutron star to be constrained to the range $8.4-15.2\,\rm km$. In this case, the maximum mass of neutron stars can be as high as $4.0 M_{\odot}$, well beyond the mass of the heaviest observed neutron stars~\cite{Demorest:2010,Antoniadis:2013}. This radius range reduces to $8.7-12.6\,\rm km$ if the EOS input is used up to twice nuclear saturation density. In this case, the upper limit on the maximum mass reduced to $2.9 M_{\odot}$. 

These results make clear that current theoretical uncertainties need to be reduced in the density range of $1-2 n_0$ in order to enable accurate theoretical predictions of neutron-star observables. Interestingly, this density range is also accessible in terrestrial heavy-ion--collision experiments, and it provides an ideal overlap for nuclear experiments, theory, and astrophysical observations.

\section{Issues of local chiral interactions and possible improvements}\label{sec:Issues}

\subsection{Fierz ambiguity}\label{sec:Fierz}

As discussed in \cref{sec:LocChi}, local chiral interactions are constructed by making use of Fierz ambiguities for short-range contact interactions in order to eliminate sources of nonlocality. At LO, the most general interaction is given in \cref{eq:LOpot}.
By using the antisymmetrization operator $\mathcal A$
\begin{align}
    \mathcal{A}\,f(\vb{q},\vb{k})=\frac14 (1+\bm\sigma_1\cdot\bm\sigma_2)(1+\bm\tau_1\cdot\bm\tau_2)\,f\left(\vb{q}\to -2\vb{k},\vb{k}\to -\frac12\vb{q} \right)\,,
\end{align}
to construct the antisymmetrized interaction
\begin{align}
    V_{\rm as}(\vb{q},\vb{k})=\frac12\Big( V(\vb{q},\vb{k})-\mathcal{A}\,V(\vb{q},\vb{k})\Big)\,,
\end{align}
one finds
\begin{align}
    V_{\rm cont, as}^{\rm LO}=\Tilde{C}_S \mathbbm{1}+\Tilde{C}_T\,\bm\sigma_1\cdot\bm\sigma_2+\left(-\frac23\Tilde{C}_S-\Tilde{C}_T  \right) \bm\tau_1\cdot\bm\tau_2 -\frac13\,\Tilde{C}_S \,\bm\sigma_1\cdot\bm\sigma_2\,\bm\tau_1\cdot\bm\tau_2\,,
    \label{eq:asym}
\end{align}
where only two LECs are linearly independent, as discussed in \cref{sec:LocChi}. 

While this reasoning has been used to choose a local set of contact operators in the derivation of local chiral interactions, when applying a regulator function it remains valid only when the regulator behaves as 
\begin{align}
    f(\vb{q},\vb{k})=f\left(-2\vb{k},-\frac12\vb{q}\right)\,.
    \label{eq:CondReg}
\end{align}
This relation is fulfilled for typical nonlocal regulators, but it can be immediately seen that local regulators do not satistfy \cref{eq:CondReg}. Therefore, when applying the antisymmetrization operator to a locally regulated interaction, one finds
\begin{align}
    V_{\rm cont, as}^{\rm LO, reg}=\Tilde{C}_S \mathbbm{1}+\Tilde{C}_T\,\bm\sigma_1\cdot\bm\sigma_2+\left(-\frac23\Tilde{C}_S-\Tilde{C}_T \right) \bm\tau_1\cdot\bm\tau_2 -\frac13\Tilde{C}_S\,\bm\sigma_1\cdot\bm\sigma_2\,\bm\tau_1\cdot\bm\tau_2+V_{\rm corr}^f(\vb{p}\cdot\vb{p}')\,,
    \label{eq:asymwithReg}
\end{align}
where the correction term depends on the form of the local regulator function~\cite{Huth:2017}. This correction is a simple manifestation of the fact that a regulator function affects potential terms beyond the order at which one is working. This effect will be corrected when including contact operators at higher orders. In fact, even a regulator which fulfills \cref{eq:CondReg} appears as a global factor in front of \cref{eq:asym}, therefore introducing higher-order correction terms $\propto p^{2n}$ and $\propto p'^{2n}$. Local regulators, instead, introduce regulator artifacts $\propto \vb{p}\cdot \vb{p}'$ and, therefore, lead to a mixing of different partial waves. For instance, the correction in \cref{eq:asymwithReg} mixes LO $S$-wave contact interactions into $P$-waves, which nonlocal regulators do not do~\cite{Huth:2017}. This effect has been found to lead to sizable contributions in the 3N sector; see Sec.~\ref{sec:results} and Refs.~\cite{Lynn:2016,Dyhdalo:2016ygz}.

\begin{figure}[htb]
\centering
\includegraphics[width=0.495\textwidth]{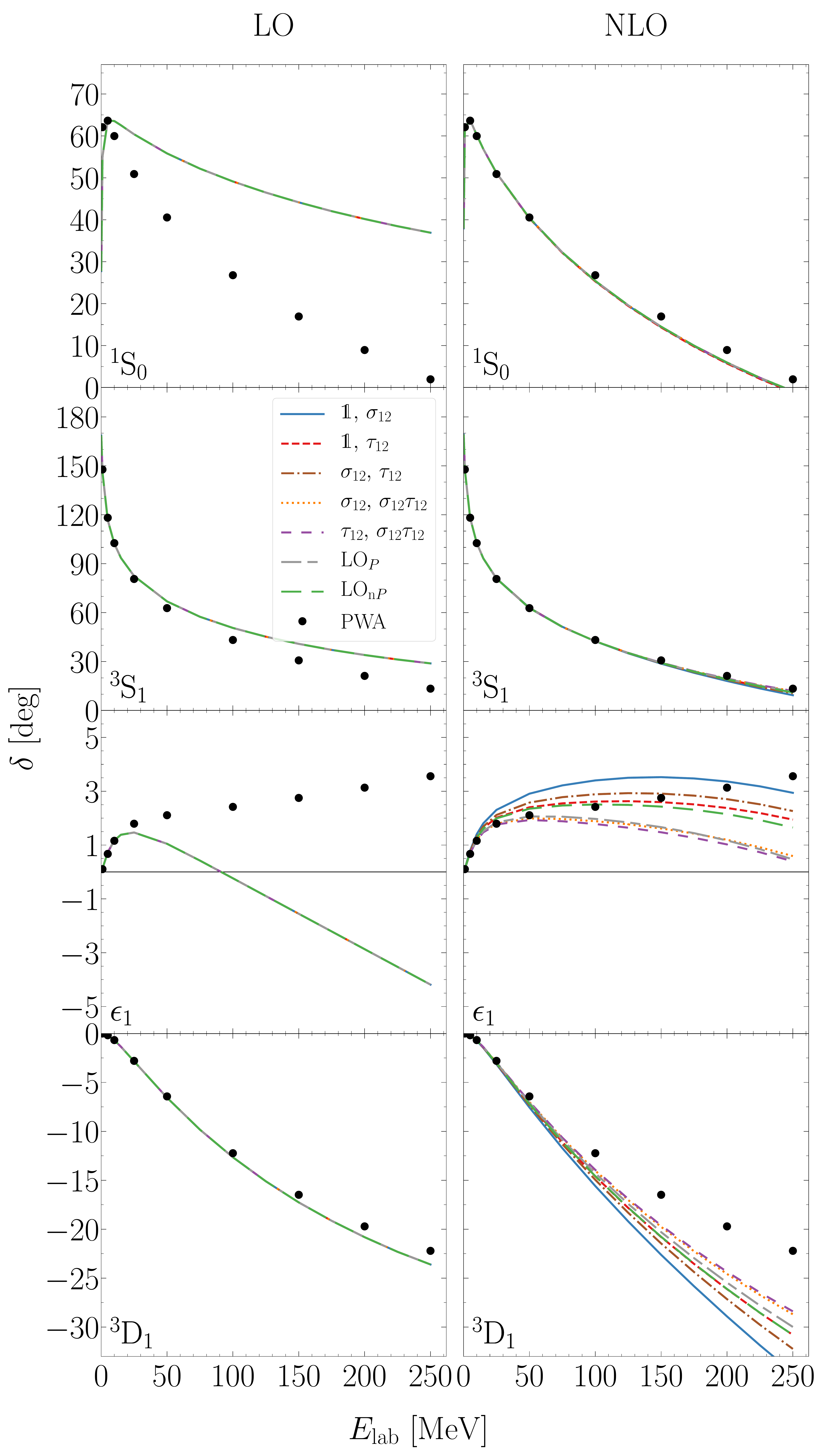}
\includegraphics[width=0.495\textwidth]{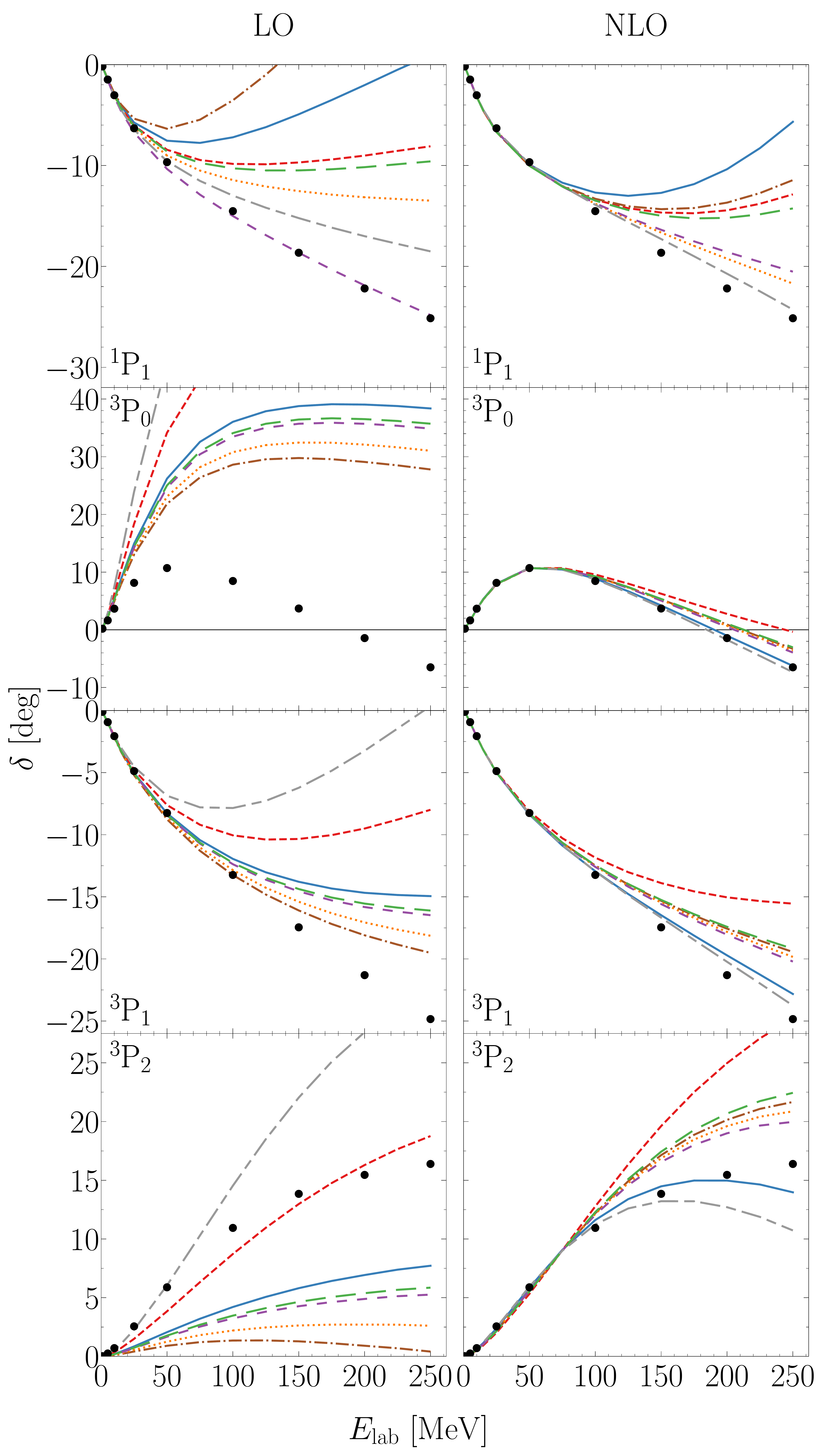}
\caption[]{$S$-wave and $P$-wave phase shifts at LO and NLO for chiral interactions with all possible LO operator combinations and $R_0=1.0\,\rm fm$~\cite{Huth:2017}.}
\label{fig:PSFierz}
\end{figure}

In Ref.~\cite{Huth:2017}, we have investigated the violation of Fierz rearrangement freedom at LO in the chiral expansion. In particular, we have constructed LO and NLO interactions for all allowed choices of operator pairs in \cref{eq:LOpot} by fitting them to $S$- and $P$-wave phase shifts. In addition, we have investigated interactions including all four LO operators, where the additional two LECs were determined so that the regulator artifacts vanish in the $P$-waves (LO$_{nP}$), or by fitting them to the $^1P_1$ and $^3P_2$ partial waves (LO$_P$). We show the resulting phase shifts in \cref{fig:PSFierz}.

At LO, while the $^1S_0$ and $^3S_1-^3D_1$ coupled channel are reproduced similarly for all operator combinations, the $P$-wave phase shifts strongly depend on the operator choice. This uncertainty was found to be larger than typical truncation uncertainty estimates~\cite{Huth:2017}, which has a significant impact on the description of many-body systems. For instance, Ref.~\cite{Huth:2017} found the EOS of PNM at LO to dramatically depend on the operator choice, with some operators even leading to bound neutron matter at nuclear saturation density. Going to NLO and including the first correction terms was found to restore the Fierz ambiguity to a large extent, and to improve the $P$-wave phase shifts as well as many-body properties; see \cref{fig:PSFierz} and Ref.~\cite{Huth:2017}. 

Similarly to the NN sector, this behavior also persists in the 3N sector, as shown in \cref{sec:results} for nuclei and neutron matter. The leading 3N contact interaction $V_E$ depends on a general operator set, similarly to NN contact interactions, and is given by~\cite{Epelbaum:2002vt}
\begin{align}
    V_E\propto c_E \sum_{i<j<k}\,\sum_{\text{cyc}}\,\mathcal{O}_{ijk}\,f_{\text{short}}(r_{ij})\,f_{\text{short}}(r_{kj})\,,
\end{align}
with
\begin{align}
    \mathcal{O}_{ijk}\in \Big\lbrace \mathbbm{1}, \bm\sigma_i\cdot\bm\sigma_j, \bm\tau_i\cdot\bm\tau_j, \bm\sigma_i\cdot\bm\sigma_j\, \bm\tau_i\cdot\bm\tau_j, \bm\sigma_i\cdot\bm\sigma_j\, \bm\tau_i\cdot\bm\tau_k, \big[(\bm\sigma_i\times\bm\sigma_j)\cdot\bm\sigma_k\big] \big[(\bm\tau_i\times\bm\tau_j)\cdot\bm\tau_k\big] \Big\rbrace\,.
\end{align}
While, once again, the choice of the operators should not influence the final result, for local regulators the Fierz rearrangement freedom is violated, and different 3N operator choices affect the predictions for nuclear systems, see \cref{sec:results}. In particular, we have explored the predictions for PNM when choosing $\mathcal{O}_{ijk}= \bm \tau_i\cdot \bm \tau_j$ ($E\tau$), $\mathcal{O}_{ijk}=\mathbbm{1}$ ($E\mathbbm{1}$), and for a projector on triples with $S=1/2$ and $T=1/2$ ($EP$). The impact of these different choices is particularly strong for softer interactions~\cite{Lonardoni:2018prl,Lonardoni:2018prc}. A possible solution to these regulator artifacts is the inclusion of subleading contact interactions, but the subleading 3N contact terms appear only at N$^4$LO. The consistent implementation of chiral forces at N$^4$LO, however, is currently not feasible. 

\subsection{Three-nucleon two-pion-exchange interaction}

\begin{figure}[t]
\centering
\includegraphics[width=0.45\textwidth]{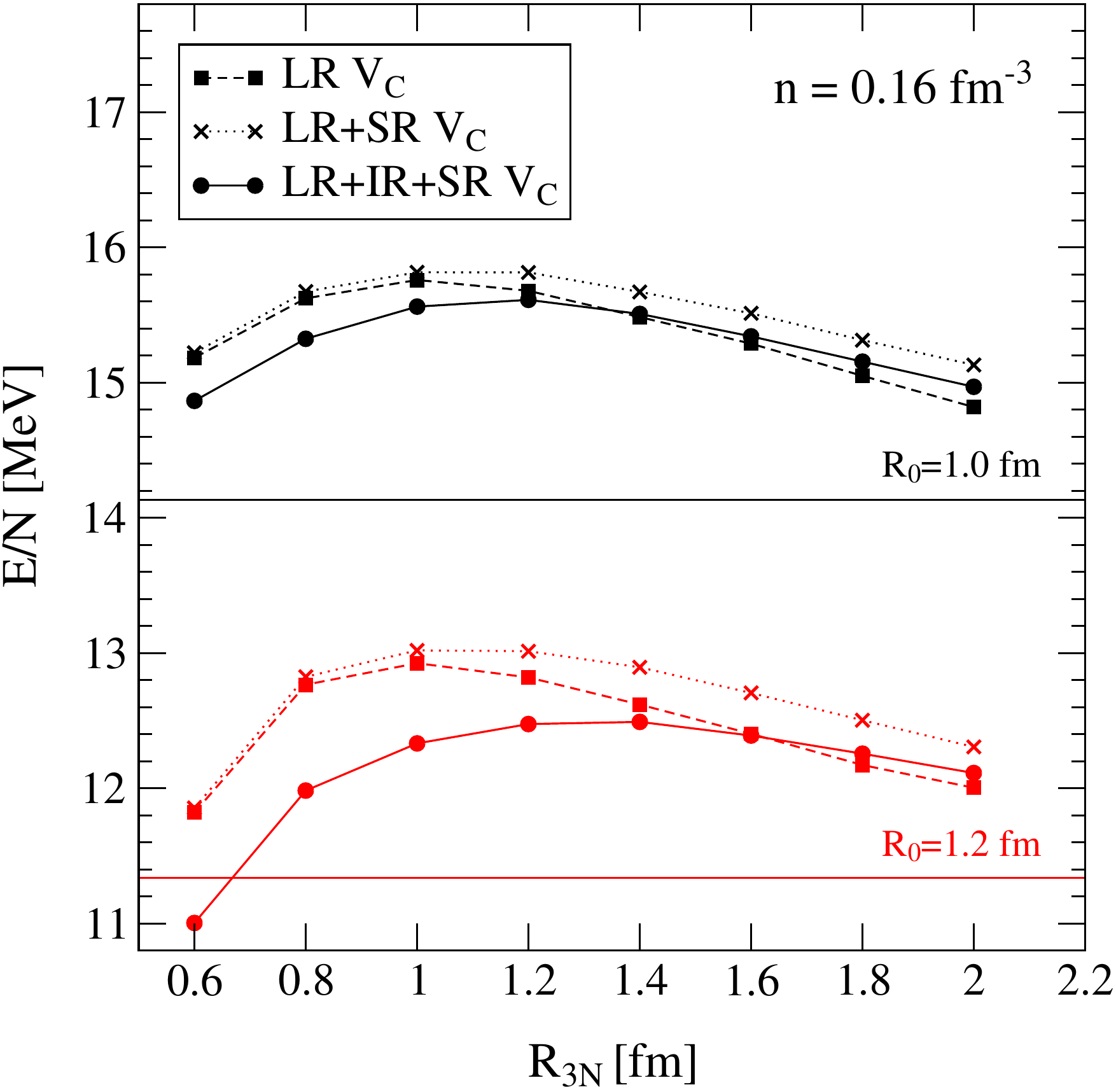}\hspace*{1cm}
\includegraphics[width=0.435\textwidth]{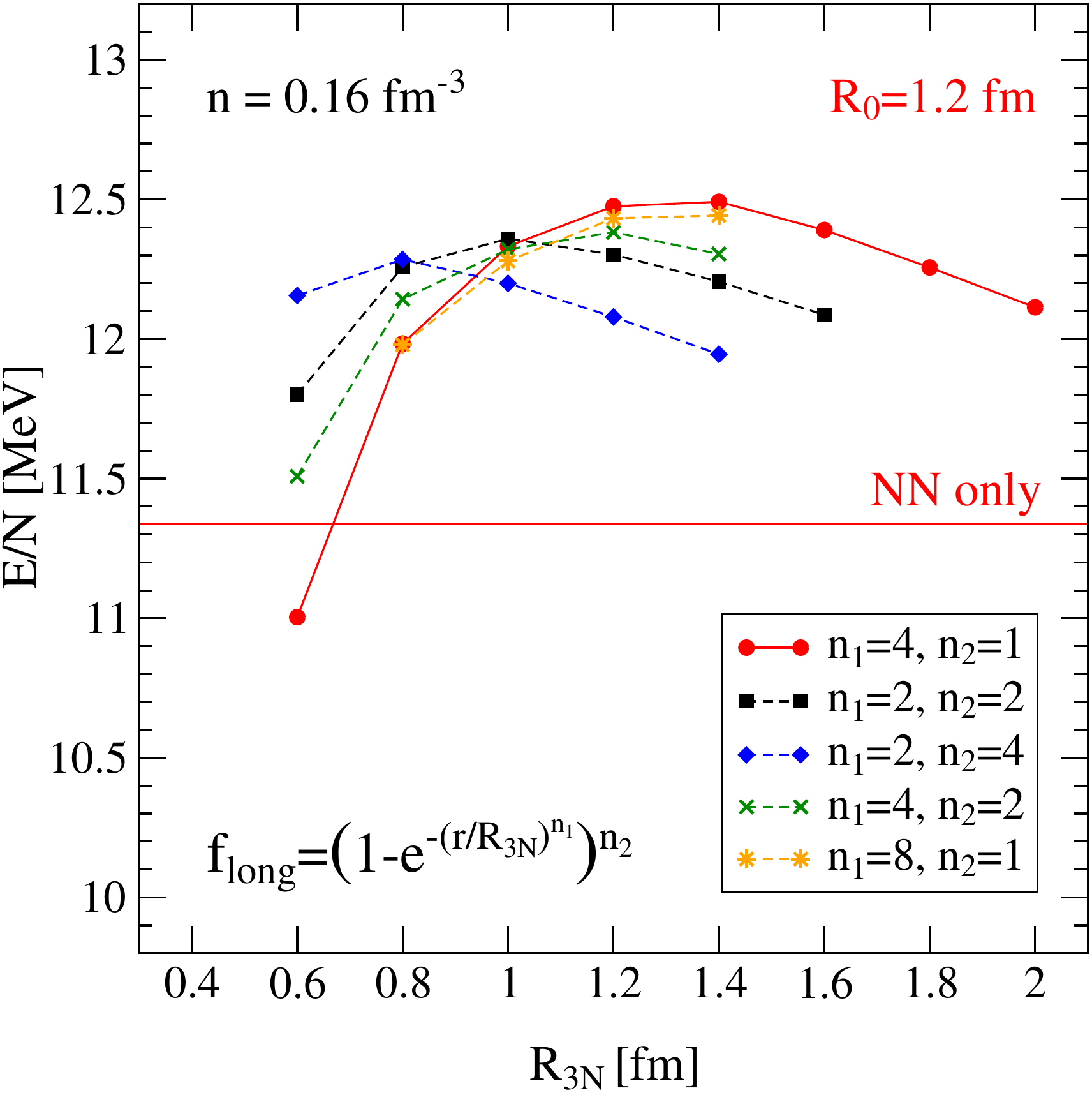}
\caption[]{Variation of the AFDMC energy per particle in PNM at nuclear saturation density as a function of the 3N cutoff $R_{3N}$~\cite{Tews:2016}. Left panel: Variation for $R_0=1.0\,\rm fm$ (black lines) and $R_0=1.2\,\rm fm$ (red lines). The horizontal lines correspond to the NN-only result. The different curves correspond to $V_C$ when including different parts of the interactions: The squares are for the long-range $c_1$ and $c_3$ terms of $V_C$, the crosses additionally include the short-range $c_3$ term of $V_C$, and the circles include all terms of $V_C$. Right panel: Variation for different exponents $n_1$ and $n_2$ in the long-range regulator function for $R_0=1.2\,\rm fm$.}
\label{fig:3NTPE}
\end{figure}

In order to study in detail the effect of the local 3N TPE interaction on the EOS of PNM, AFDMC calculations have been carried out including only this component of the 3N force~\cite{Tews:2016}. We find that a locally-regulated 3N TPE adds less repulsion to PNM than a nonlocally-regulated version. In \cref{fig:3NTPE}, we show the variation of the PNM energy per particle at saturation density as a function of the 3N cutoff $R_{3N}$ when only the TPE interaction is included. The horizontal lines represent the NN-only results. We find that the maximum contribution of the 3N TPE is found when $R_0=R_{3N}$, and that this contribution is of the order of $1-2\,\rm MeV$, only about half as large as for nonlocal 3N TPE interactions~\cite{Hebeler:2009iv}. The discussed behavior is independent of the exact regulator form, and it is not impacted by the additional shorter-range terms that appear upon Fourier transformation of the momentum-space TPE expression. The maximal 3N TPE contribution is larger for larger $R_0$. When $R_{3N}$ is significantly smaller than $R_0$, the system collapses, as the 3N attraction overcomes the NN repulsion~\cite{Tews:2016}.

This behavior has also been investigated in detail in Ref.~\cite{Dyhdalo:2016ygz}. In short, local regulators lead to a smaller effective cutoff for the 3N TPE interaction as compared to nonlocal regulators. As a result, larger 3N cutoffs are needed to reduce these regulator artifacts. Larger 3N cutoffs, in addition, will also reduce the regulator artifacts due to the violation of Fierz rearrangement freedom. However, as shown in \cref{fig:3NTPE}, larger 3N cutoffs also require larger NN cutoffs to avoid collapses. 

\subsection{Local LO interactions at large cutoffs}\label{sec:LargeCutoff}

As introduced in the previous section, a possible solution to minimize local regulator artifacts is to construct nuclear interactions with larger coordinate-space cutoffs. While such interactions are not practical for most many-body methods, as they are typically too hard to lead to a reasonable many-body convergence, QMC methods can efficiently treat harder interactions without difficulties. 

\begin{figure}[t]
\centering
\includegraphics[width=0.99\textwidth]{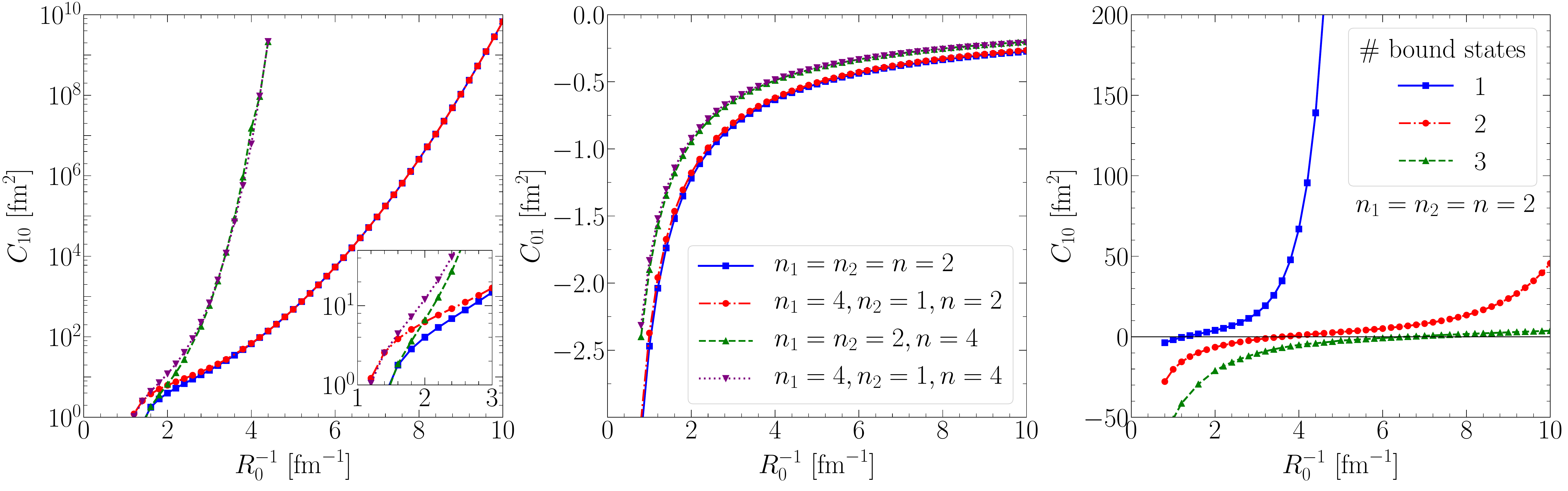}\hspace*{1cm}
\caption[]{Spin-isospin LECs $C_{ST}$, $C_{10}$ (left panel, logarithmic scale) and $C_{01}$ (central panel), as functions of the inverse cutoff $R_0^{-1}$ for local chiral interactions at LO with different local regulators characterized by $n_1$, $n_2$, and $n$ exponents~\cite{Tews:2018sbi}. Right panel: Spin-isospin LEC $C_{10}$ as function of the inverse cutoff $R_0^{−1}$ for local chiral interactions at LO enforcing one, two, and three bound states in the coupled $^3S_1-^3D_1$ channel.}
\label{fig:LOLecsHighCut}
\end{figure}

To explore this possibility, we have constructed local chiral interactions at LO over a wide cutoff range~\cite{Tews:2018sbi}. In particular, we have constructed potentials for all possible LO operator structures with corresponding momentum-space cutoffs ranging from $400$ to $4000\,\rm MeV$. When fitting to phase shifts, for smooth local regulators there is an ambiguity in the number of bound states in the $^3S_1$ channel. An arbitrary number of bound states can be realized; see the right panel of \cref{fig:LOLecsHighCut}. For the different number of bound states, the spectral LEC $C_{10}$, where the indices denote $S$ snd $T$, respectively, dramatically varies in size.

When fitting our interactions with smooth local regulators, it is possible to enforce exactly one bound state in the coupled $^3S_1-^3D_1$ channel. While, typically, spurious bound states appear in partial waves with attractive tensor interactions when the cutoff is increased, enforcing one bound state in the deuteron channel leads to an increasing spectral LEC $C_{10}$ in the deuteron channel. Its magnitude, together with the smoothness of the regulator function, ensures that the attractive parts of the one-pion exchange leading to spurious bound states are cut off. 

\begin{figure}[htb]
\centering
\includegraphics[width=0.99\textwidth]{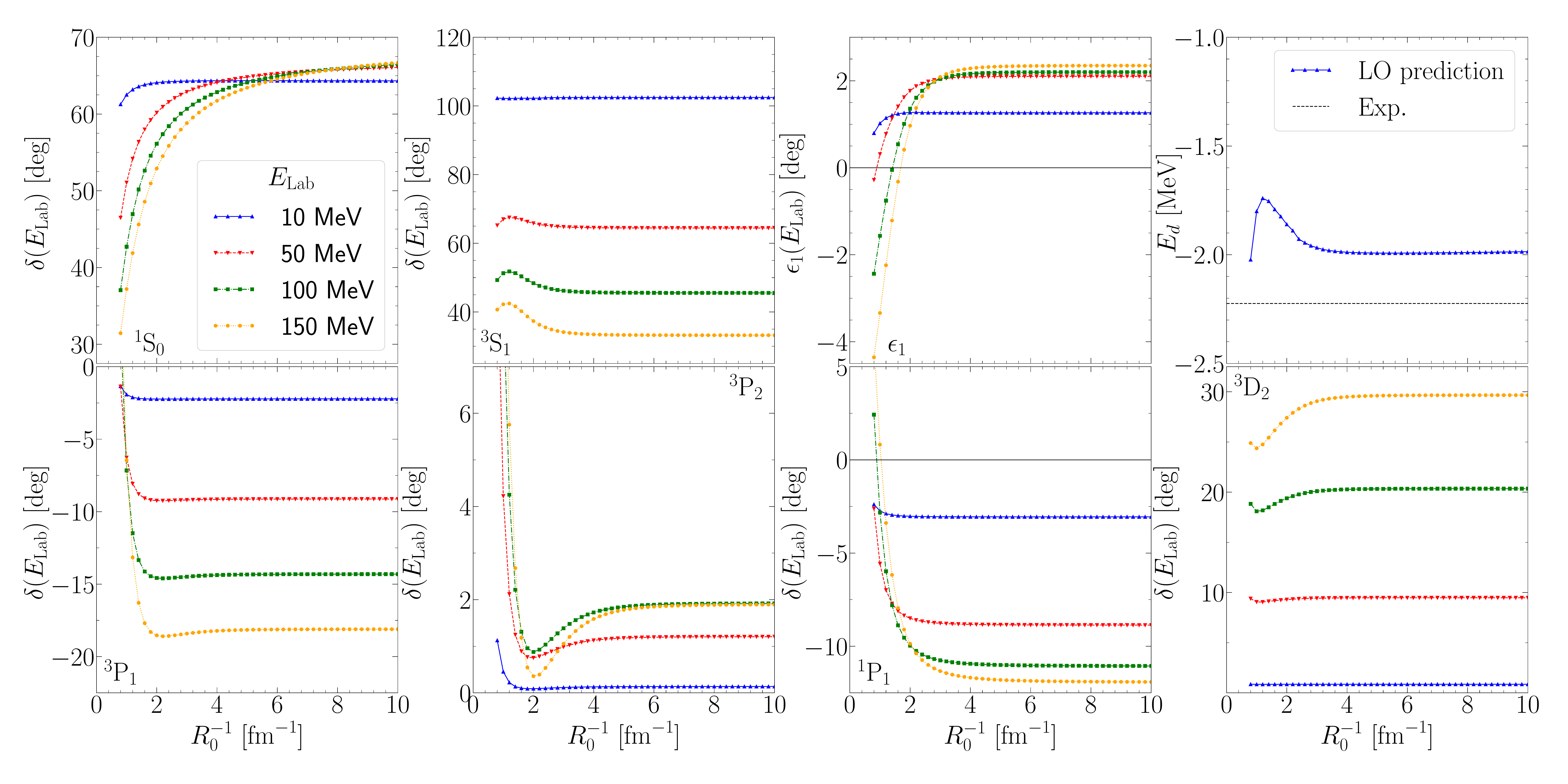}\hspace*{1cm}
\caption[]{Phase shifts in the ${}^1S_0, {}^3S_1, {}\epsilon_1, {}^3D_2, {}^1P_1,{}^3P_1$, and ${}^3P_2$ partial waves for laboratory energies $E_{\text{lab}}=10, 50, 100$, and $150\,\rm  MeV$, as well as the deuteron ground-state energy $E_d$ (upper right panel) as functions of the inverse cutoff $R_0^{-1}$ for the LO operators $\big\{\mathbbm{1}$, $\bm\sigma_1\cdot\bm\sigma_2\big\}$ and $n_1=n_2=n=2$~\cite{Tews:2018sbi}.}
\label{fig:LOhighCut}
\end{figure}

Due to the locality of the interactions, this short-range repulsion in the $^3S_1$ partial wave is also mixed into higher partial waves, but its magnitude and sign depend on the LO operator structure. We find that for the operators $\big\{\mathbbm{1}$, $\bm\sigma_1\cdot\bm\sigma_2\big\}$ the regulator artifacts are repulsive in all higher partial waves with attractive tensor contributions, where spurious bound states might appear. These regulator artifacts are also strong enough to compensate for the strong attraction from the one-pion-exchange interaction and help to avoid spurious bound states. In addition, we find a stabilization of phase shifts and the deuteron binding energy on cutoff independent plateaus, see \cref{fig:LOhighCut}. We stress that this does not imply that these interactions are renormalizable. We simply use local regulator artifacts to our advantage to construct large-cutoff chiral interactions suitable for QMC calculations. These interactions will ultimately help us to increase the 3N cutoff and to reduce the 3N regulator artifacts, and we are currently working on larger-cutoff interactions at higher orders in chiral EFT. 

\section{Summary}\label{sec:summary}

In this proceedings, we have presented recent results from QMC calculations with local chiral EFT interactions and discussed issues and possible future directions. We have shown that QMC calculations with local chiral interactions up to N$^2$LO give an excellent description of nuclei up to \isotope[16]{O}, pure neutron matter, and $n$-$\alpha$ elastic scattering phase shifts, providing interesting insights on nuclear structure, short-range correlations, and the properties of neutron stars. 

However, the employed potentials can lead to sizable local regulator artifacts that have to be carefully analyzed. In particular, the violation of the Fierz ambiguity in the 3N sector and lower effective cutoffs for the 3N TPE lead to sizable 3N regulator artifacts, that increase uncertainties for nuclei and neutron matter. While a solution would be to include subleading 3N short-range interactions, the implementation of these contribution is currently not feasible. As a possible alternative, we have discussed large-cutoff chiral interactions that can be treated with QMC methods and also allow to increase the 3N cutoff, leading to smaller cutoff artifacts. We have presented some results for such local interactions at LO, and we are currently exploring higher-order potentials. 

\acknowledgments
We would like to thank the organizers for the invitation and for delivering a successful workshop. A special thank goes to J. Carlson, J.-W. Chen, W. Detmold, E. Epelbaum, S. Gandolfi, A. Gezerlis, K. Hebeler, L. Huth, O. Hen, A. Lovato, J. E. Lynn, J. Margueron, A. Nogga, C. Petrie, S. Reddy, K. E. Schmidt, A. Schwenk, X. B. Wang, R. B. Wiringa, for insightful discussions and their contribution to the studies presented in this work. This work was supported by the U.S. Department of Energy, Office of Science, Office of Nuclear Physics, under Contracts No.~DE-SC0013617 and~DE-AC52-06NA25396, and under the FRIB Theory Alliance award DE-SC0013617, by the Los Alamos National Laboratory (LANL) LDRD program, and by the NUCLEI SciDAC program. This research used resources provided by the LANL Institutional Computing Program, which is supported by the U.S. Department of Energy National Nuclear Security Administration under Contract No.~89233218CNA000001. Computational resources have also been provided by the National Energy Research Scientific Computing Center (NERSC), which is supported by the U.S. Department of Energy, Office of Science, under Contract No. DE-AC02-05CH11231, and by the J\"ulich Supercomputing Center.


\providecommand{\href}[2]{#2}\begingroup\raggedright\endgroup

\end{document}